\documentclass[review]{elsarticle}

\usepackage{subcaption}

\usepackage{lineno,hyperref}
\usepackage{multirow}
\usepackage{rotating}

\usepackage{amsmath}
\usepackage{cases}

\usepackage{tabu}
\usepackage{booktabs}

\usepackage{xcolor}

\usepackage{pifont}
\newcommand{\cmark}{\ding{51}}
\newcommand{\xmark}{\ding{55}}

\usepackage{glossaries}
\usepackage{glossary-mcols}
\glsdisablehyper
\makeglossaries
\newacronym{IoT}{IoT}{Internet of Things}
\newacronym{QoS}{QoS}{Quality of Service}
\newacronym{RTT}{RTT}{Round-Trip Time}
\newacronym{MS}{M\&S}{Modeling and Simulation}
\newacronym{MBSE}{MBSE}{Model-Based Systems Engineering}
\newacronym{DUNIP}{DUNIP}{DEVS Unified Process}
\newacronym{DEVS}{DEVS}{Discrete Event System Specification}
\newacronym{DEVSML}{DEVSML}{DEVS Modeling Language}
\newacronym{ADAS}{ADAS}{Advanced Driver Assistance System}
\newacronym{ISP}{ISP}{Internet Service Provider}
\newacronym{RAN}{RAN}{Radio Access Network}
\newacronym{UE}{UE}{User Equipment}
\newacronym{AP}{AP}{Access Point}
\newacronym{EDC}{EDC}{Edge Data Center}
\newacronym{NR}{NR}{5G New Radio}
\newacronym{PBCH}{PBCH}{Physical Broadcast Channel}
\newacronym{PSS}{PSS}{Primary Synchronization Signal}
\newacronym{SNR}{SNR}{Signal-to-Noise Ratio}
\newacronym{PRACH}{PRACH}{Physical Random Access Channel}
\newacronym{AMF}{AMF}{Access and Mobility Management Function}
\newacronym{FDD}{FDD}{Frequency Division Duplexing}
\newacronym{PUCCH}{PUCCH}{Physical Uplink Control Channel}
\newacronym{PDCCH}{PDCCH}{Physical Downlink Control Channel}
\newacronym{PUSCH}{PUSCH}{Physical Uplink Shared Channel}
\newacronym{PDSCH}{PDSCH}{Physical Downlink Shared Channel}
\newacronym{MCS}{MCS}{Modulation and Codification Scheme}
\newacronym{RRC}{RRC}{Radio Resource Control}
\newacronym{SysML}{SysML}{Systems Modeling Language}
\newacronym{GPU}{GPU}{Graphics Processing Unit}
\newacronym{CPU}{CPU}{Central Processing Unit}
\newacronym{FSM}{FSM}{Finite State Machine}
\newacronym{DVFS}{DVFS}{Dynamic Voltage and Frequency Scaling}
\newacronym{SDN}{SDN}{Software-Defined Network}
\newacronym{ANN}{ANN}{Artificial Neural Network}
\newacronym{NRMSD}{NRMSD}{Normalized Root Mean Square Deviation}
\newacronym{GPS}{GPS}{Global Positioning System}
\newacronym{EMA}{EMA}{Exponential Moving Average}
\newacronym{QAM}{QAM}{Quadrature Amplitude Modulation}
\newacronym{WLAN}{WLAN}{Wireless Local Area Network}
\newacronym[plural=ABC,firstplural=Abstract Base Classes (ABCs)]{ABC}{ABC}{Abstract Base Class}
\newacronym{DNN}{DNN}{Deep Neural Network}
\newacronym{CNN}{CNN}{Convolutional Neural Network}
\newacronym{CSV}{CSV}{Comma-Separated Values}
\newacronym{M2M}{M2M}{Machine-to-Machine}
\newacronym{LTE}{LTE}{Long Term Evolution}
\newacronym{WAN}{WAN}{Wide Area Network}
\newacronym{V2V}{V2V}{Vehicle-to-Vehicle}
\newacronym{V2I}{V2I}{Vehicle-to-Infrastructure}
\newacronym{MSO}{M\&S\&O}{Modeling, Simulation, and Optimization}
\newacronym{FaaS}{FaaS}{Function as a Service}
\newacronym{MAN}{MAN}{Metropolitan Area Network}
\newacronym{IBD}{IBD}{Internal Block Description}
\newacronym{STM}{STM}{State Machine}

\usepackage{url}
\usepackage{breakurl}

\usepackage[htt]{hyphenat}

\journal{Journal of Simulation Modelling Practice and Theory}

\begin{document}

\begin{frontmatter}

\title{Mercury: a Modeling, Simulation, and Optimization Framework for Data Stream-Oriented IoT Applications}


%
%

\author[lsi]{Rom\'{a}n~C\'{a}rdenas\corref{cor1}}
\ead{r.cardenas@upm.es}
\author[lsi,ccs]{Patricia~Arroba}
\ead{parroba@die.upm.es}
\author[lsi]{Roberto~Blanco}
\ead{r.bandres@die.upm.es}
\author[lsi,ccs]{Pedro~Malag\'{o}n}
\ead{malagon@die.upm.es}
\author[ucm,ccs]{Jos\'{e}~L.~Risco-Mart\'{i}n}
\ead{jlrisco@ucm.es}
\author[lsi,ccs]{Jos\'{e}~M. Moya}
\ead{jm.moya@upm.es}

\cortext[cor1]{Corresponding author}
\address[lsi]{Laboratorio de Sistemas Integrados, Universidad Polit\'{e}cnica de Madrid, 28040, Spain}
\address[ucm]{Dpt. of Computer Architecture and Automation, Universidad Complutense de Madrid, Madrid 28040, Spain}
\address[ccs]{Center for Computational Simulation, Universidad Polit\'{e}cnica de Madrid, 28660, Spain}

\begin{abstract}
The Internet of Things is transforming our society by monitoring users and
infrastructures' behavior to enable new services that will
improve life quality and resource management. These applications require a vast
amount of localized information to be processed in real-time so, the deployment
of new fog computing infrastructures that bring computing closer to the data
sources is a major concern.  In this context, we present Mercury, a Modeling, Simulation, and Optimization (M\&S\&O)
framework to analyze the dimensioning and the dynamic operation of real-time
fog computing scenarios. Our research proposes a location-aware solution that
supports data stream analytics applications including FaaS-based computation
offloading. Mercury
implements a detailed structural and behavioral simulation model, providing fine-grained simulation outputs, and is described using the
Discrete Event System Specification (DEVS) mathematical formalism, helping to validate the model's implementation.
Finally, we present a case study using real traces from a driver assistance
scenario, offering a detailed comparison with other state-of-the-art simulators.

\end{abstract}

\begin{keyword}
Fog Computing\sep Edge Federation\sep MBSE\sep IoT\sep Data Stream\sep 5G
\end{keyword}

\end{frontmatter}

\section{Introduction}
\label{sec:1_intro}
The escalating process of digitalization is triggering the proliferation of new
\gls{IoT} services and applications. The \gls{IoT} technological revolution is
transforming our society by recording and analyzing the behavior of users and
infrastructures to develop services for improving quality of life and resource
management. This technology is in full growth, and the deployment of tens of
billions of devices connected to the internet is expected within the next few
years~\cite{GartnerPR2017}. 

Computing \gls{IoT}-based environments need Big Data infrastructures for
effective real-time decision making, capable of processing large amounts of
complex data from multiple, geographically distributed sources. Moreover, many
\gls{IoT} applications demand low latencies and rapid mobility~\cite{deng2016}.
However, for Big Data when response time is critical, cloud computing is not
capable of meeting these needs and traditional networks  would also impose
overloads to the communication bandwidth of the core
network~\cite{IECedge2017}.

Therefore, a new paradigm is emerging. Fog computing was introduced by
researchers from Cisco Systems in 2012~\cite{cisco2012}. It extends the cloud
computing paradigm to the edge of the network, deploying micro data centers
that operate on the edge and provides a drastic latency reduction compared to
cloud computing~\cite{shi2016}.  These \glspl{EDC} process information from
heterogeneous \gls{IoT} data sources, perform the operations that require the
minimum response time, and aggregate the data for subsequent sending to the
cloud, thus decongesting the communications network and avoiding its
saturation~\cite{8089336}.  Recent research reports that in this computing
paradigm the \gls{RTT} may be reduced by 92\% and the overall available
bandwidth may be increased by a factor of 47-58 when compared to a cloud
platform~\cite{yi2015}.

Many relevant \gls{IoT} applications can take advantage of these features, as
i) those based on distributed monitoring systems using embedded devices with
limited processing capacity (e.g., environmental analysis), and ii) real-time
data stream analytics, processing a large volume of data (e.g., e-health and
driving assistance systems)~\cite{ansys2017,8289317}.  However, the new edge
infrastructure needed to support fog computing requires the deployment of
numerous smaller data centers that can be relocated to bring computing closer
to the origin of the information. Moreover, those data centers' energy
consumption must be within the margins supported by the electricity grid. To
reduce the consumption of the edge infrastructure and improve its availability,
the concept of an edge federation arises. 

The federation of \glspl{EDC} allows the combined management of both the
resource demand depending on the volume of users, their physical location and
the use of the available renewable energy sources, thus reducing the carbon
footprint~\cite{8360857}.  Also, this location-awareness particularities of fog
computing make the  \gls{FaaS}-based management of applications of great
interest to further optimize resource utilization in the edge
federation~\cite{glikson2017}.

Moreover, \gls{IoT} applications depict a pervasive network of multiple
distributed wireless devices that are both resource and energy-limited, all
able to collect and send large volumes of heterogeneous environmental
information in real-time. Due to the
energy-computing-bandwidth limitations of the wireless domain, the 5G mobile communication standard arises as a
service enabler to overcome these technical barriers. 5G appears as a highly
flexible technology that provides a solid field for \gls{IoT} applications with diverse requirements
deployment. Together with \gls{IoT} and fog computing, 5G promises to cause a
technology disruption in a wide range of business sectors.

Currently, \gls{IoT} models and simulators are in great need, since the
conception, design, implementation, and deployment of such complex
architectures as edge federations, demand in-depth virtual analysis.  The
nature of the problem suggests the application of solid advanced \gls{MS}
methods based on the principles of \gls{MBSE} that ensures a logic, robust, and
reliable incremental design. 
We capitalize on the
principles of \gls{MBSE} of evolving the model - defined using the \gls{DEVS}
formalism - as the central artifact throughout the lifecycle of the system. 

In this context we present Mercury, a \gls{MSO} framework designed to explore
the dimensioning and the dynamic operation of real-time fog computing
scenarios. Mercury is open source, and it can be downloaded from its official
repository~\cite{mercuryrepo}.  Particularly, the contributions of our research
are the following:
\begin{itemize}
\item Our model exploits the location-awareness inherent in fog computing by
implementing a FaaS-based computation offloading over a network with 5G
capabilities. Mercury supports data stream analytics applications, as they may
obtain the most benefits from this scenario, as real-time low latency and high
data volume throughput, avoiding network congestion.
\item We provide a high-detailed structural and behavioral simulation model
based on \gls{MBSE}, where the DEVS mathematical formalism helps to validate
the atomic models’ implementation.
\item This approach allows Mercury to provide more detailed simulation outputs,
including real-time data for heterogeneously configured IT and network
resources. We also provide an allocation manager to automatically optimize the
configuration of the model to be explored. This makes Mercury a powerful
\gls{MSO} framework to analyze both location-aware deployments and dynamic
operation strategies for the edge infrastructure.
\item In this paper, as a case study, we model a realistic \gls{ADAS}
application based on data stream analytics using real traces from traffic
mobility and from vehicles’ onboard cameras. Also, we offer a detailed
comparison with other state-of-the-art fog computing simulators to highlight
Mercury's novelties.
\end{itemize}

The present research paper is a significant extension of our previous
research~\cite{acceptedSummersim}, in which we presented an edge
federation \gls{MS} framework for data stream analytics focused on evaluating the
power and delay perceived at the edge layer. In the present publication we
significantly extend this approach to introduce Mercury. Following are the main contributions
added to our previous research:
\begin{itemize}
\item Mercury now includes a detailed 5G-based model (e.g., radio interface,
physical interface, bandwidth assignation schematics, signal encoding and bit
rates) to increase the realism of the radio interfaces’ impact on fog
computing. Hence, the new research presents a solution for a complete fog scenario
instead of focusing only on the edge computing infrastructure.
\item Mercury allows the user to monitor, not only power and delay, but also
bandwidth share, spectral efficiency and bit rate at the radio interface side.
Also, 2D Mobility is currently supported natively by Mercury.
\item Our previous research provided an
\gls{MS} tool, where the placement of network and computing infrastructures
must be provided manually. Mercury is an \gls{MSO} framework that also provides
an allocation manager module for scenario optimizations, automatically
determining the location of both the network and the computing infrastructures
according to the IoT application under study. 
\item The present research has a completely novel performance evaluation.
Mercury has been validated under a realistic scenario with further complexity,
including real traces of IoT devices’ GPS location history. We provide an
evaluation of power, delay, bandwidth share, spectral efficiency and bit rate
outputs for the use-case scenario configurations under discussion.
This paper also includes a detailed performance comparison with other
state-of-the-art fog computing \gls{MS} frameworks.  
\end{itemize}
The remainder of this paper is organized as follows: in
Section~\ref{sec:2_soa}, the state-of-the-art on simulation tools for fog
computing environments is provided.  Sections~\ref{sec:3_architecture}
and~\ref{sec:4_implementation} give further information on the architecture of
Mercury and its implementation.  Section~\ref{sec:5_use_case} presents an
\gls{ADAS} use case scenario and extensively explains the performance
evaluation and a comparison with other state-of-the-art alternatives.  Finally,
in Section~\ref{sec:6_conclusions}, the main conclusions and future directions
are outlined.

\section{State of the Art}
\label{sec:2_soa}
This section presents several \gls{MS} tools that are extensively used for specific
use cases that could benefit from edge and fog computing scenarios. These frameworks
tend to focus on the potential benefits that fog computing could contribute to the
service instead of the edge computing infrastructure itself. As an example, one
widespread use case is \gls{ADAS}. Research provided by He et al.~\cite{HeF0WL18}
includes software for the recognition and location of traffic signals for driving
assistance scenarios that make use of edge computing features. In this research, three
main elements are identified: intelligent vehicles, edge servers, and cloud servers.
The functionality of each of them is detailed, as well as their interactions. However,
communication, power consumption, and latency analysis are not explained in detail.
Yuan et al.~\cite{8270636} present a content caching strategy for driving assistance
applications in which the content delivery is edge-assisted, and both \gls{V2V} and
\gls{V2I} communications are extensively used for enhancing driving assistance
functions. They also present the prediction of the service content demand in a
delay-constrained scenario. Moreover, Li et al.~\cite{LI2018667} provide an \gls{IoT}
scenario that includes both edge and cloud infrastructures for \gls{ADAS} data stream
applications. Their research presents power models and analysis for both \gls{LTE}
network and \gls{IoT} devices, but the latency analysis is not explained in detail.

Even though the previous \gls{ADAS}-based research works introduce important aspects
for service-specific purposes, they provide neither power consumption nor resource
management of the edge data centers' layer. The 5G model is not included in any of
these research works as well.

On the other hand, there are fog computing frameworks that focus on the required
infrastructure for providing fog services, regardless of the \gls{IoT} application.
These frameworks usually focus on infrastructure design parameters and are intended to
be used for fog computing infrastructure dimensioning and management.
EdgeCloudSim~\cite{edgecloudsim} presents a framework for the performance evaluation
of edge architectures. Their network model includes \glspl{MAN}, \glspl{WAN}, and
\glspl{WLAN}, but they do not provide the power consumption analysis in the
infrastructure. Gupta et al.~\cite{ifogsim} present the iFogSim simulator for modeling
and simulation of the resource management of both fog and edge scenarios. Their
research does not include location capabilities, but its development would be included
in their future directions. However, none of these edge frameworks implement a 5G-based
radio interface model nor scenario optimization capabilities. Table~\ref{tab:soa} shows
a comparison of the previously described frameworks.

\begin{table} [htb]
	\footnotesize
	\centering
	\caption {Edge computing simulators comparison.}
	\begin{tabular}{lcccccc}
  		\hline
		Research		& \cite{HeF0WL18}	& \cite{8270636}	& \cite{LI2018667}		& \cite{edgecloudsim}	& \cite{ifogsim}	& Mercury	\\
  		\hline
		Edge framework	& \cmark 		& \cmark 		& \cmark 		& \cmark 		& \cmark 	& \cmark	\\
	Generic use cases	& \xmark 		& \xmark 		& \xmark 		& \cmark 		& \cmark 	& \cmark	\\
	Mobility support	& \cmark 		& \cmark 		& \xmark 		& \cmark 		& \xmark 	& \cmark	\\
		Real time		& \cmark 		& \cmark 		& \cmark 		& \cmark 		& \cmark 	& \cmark	\\
		Data stream 	& \xmark 		& \xmark 		& \cmark 		& \xmark 		& \xmark 	& \cmark	\\
		5G model		& \textbf{-} 	& \xmark 		& \xmark 		& \xmark 		& \xmark 	& \cmark	\\
Edge power consumption	& \xmark 		& \xmark 		& \xmark 		& \xmark 		& \cmark 	& \cmark	\\
Edge optimization		& \xmark 		& \xmark 		& \xmark 		& \xmark 		& \xmark 	& \cmark	\\
Latency analysis		& \textbf{-} 	& \xmark 		& \textbf{-} 	& \cmark 		& \cmark 	& \cmark	\\ 
DC resource management	& \xmark 		& \xmark 		& \xmark 		& \cmark 		& \cmark 	& \cmark	\\
		\hline
	\end{tabular}
	\label{tab:soa}
\end{table}

In our previous research~\cite{acceptedSummersim}, we presented an
edge federation \gls{MS} framework for data stream analytics that is
focused on the power and delay perceived at the edge layer. In the
present paper, we significantly extend this approach to present
Mercury, an enhanced \gls{MSO} framework for fog computing scenarios.
Mercury includes a complete 5G-based model (e.g., radio and physical
interfaces, bandwidth assignation and signal encoding) to increase the realism of the radio's
impact on fog computing. Mercury allows the monitoring of not only
power and delay, but also bandwidth share, spectral efficiency and
bit rate at the radio interface's side. We also add an allocation
manager module for scenario optimizations, automatically determining
the location of the network and computing infrastructures according
to the \gls{IoT} application under study. Mobility is natively
supported by Mercury, allowing the simulation of more realistic
scenarios with further complexity, including real traces of \gls{IoT}
devices’ \gls{GPS} location history.

\section{Mercury Architecture Overview}
\label{sec:3_architecture}
\subsection{Fog Model}

Mercury's model states that the fog computing scenario would take place mostly within a
single \gls{RAN} of an \gls{ISP}. However, access control and network management-related
functionalities  would take place within the \gls{ISP}'s core network.

\begin{figure}[b]
	\center
	\includegraphics[width=0.85\textwidth]{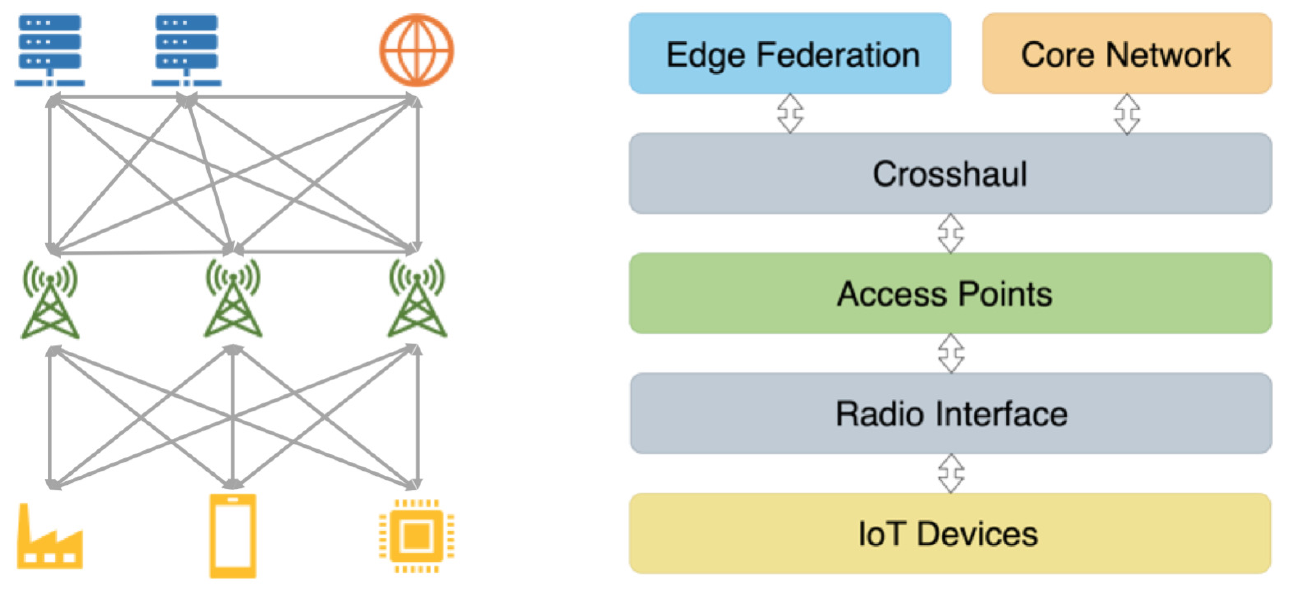}
	\caption{Mercury fog computing model.}
	\label{fig:model_scenario}
\end{figure}

Even though cloud computing is included in the concept of fog computing, the aim of Mercury
is exploring the location awareness and mobility effects on \gls{QoS} and operational
expenses for optimizing the dimensioning and operation tasks of the edge infrastructure
required to enable services that make extensive usage of computation offloading. Thus,
cloud computing is not included in Mercury's first approach of the scenario under study.
Figure~\ref{fig:model_scenario} shows an overview of the proposed scenario.
In general terms, Mercury's fog model comprises six submodules or \textit{layers}. Each
layer models different network elements that compose the fog computing scenario. 

The \textbf{\gls{IoT} Devices layer} contains all the \gls{UE} devices of the scenario. 
\gls{UE} devices are mobile, and hence their position may change over time. 
\gls{UE} may run one or more applications, and due to hardware limitations - e.g.,
computational power -, they require the transfer of resource-intensive tasks to an
external computing platform.
Mercury is primarily thought for \textit{data stream-oriented \gls{IoT} applications}, as
these could benefit the most from fog computing. Data stream-oriented \gls{IoT}
applications are defined as applications that produce large amounts of data that have to be
processed computation-intensively in real-time - e.g., video streaming~\cite{6253581}. 

Computation offloading is hosted on the \textbf{Edge Federation layer}. It is comprised of
several \glspl{EDC} located within the \gls{RAN} under study. These \glspl{EDC} contain
multiple processing units, which are the hardware that provide the processing capacity to their
corresponding \gls{EDC}. 
For modeling how computation offloading is performed in Mercury, we based our model on the
\gls{FaaS} concept, also known as serverless~\cite{serverlesscon}. In \gls{FaaS}
environments, there are not resources reserved for any application; the infrastructure will start a process only when a client makes
a service request. Due to 
location-awareness particularities of fog computing, implementing \gls{FaaS}-based
applications may optimize resource pooling and elasticity, security, provisioning, and
management at scale~\cite{glikson2017}.

In Mercury, a \gls{UE} may implement one or more applications. Each application generates
a data stream. We refer to a \textit{service} as the computation offloading
for processing the data stream of an individual \gls{UE} application. When a \gls{UE} starts a service, it sends a \textit{create session request} to one of the
\glspl{EDC} of the edge federation. A \textit{session} consists of reserving computing
resources temporarily for the exclusive processing of data flows from a given service. In this way, enough computing resources are granted until the \gls{UE}
terminates the session. Once a session is closed, computing resources are released and
available for any other potential new service. 

\gls{UE} connect to the \gls{RAN} via \glspl{AP} scattered around the scenario. These
\glspl{AP} are gathered in the \textbf{Access Points layer}. \glspl{AP} also manage the \textbf{Radio Interface layer}'s resources. This layer interconnects
\gls{UE} with \glspl{AP}.
The \textbf{Core Network layer} gathers the functions that are to be performed within the
\gls{ISP}'s core network. These functions are enforcing access control policies in the
\gls{RAN} and configuring the \textbf{Crosshaul layer} interconnections. The Crosshaul
layer interconnects \glspl{EDC}, \glspl{AP} and the core network functions.


\subsection{Allocation Manager}
\begin{figure}[htb]
        \center
        \includegraphics[width=\textwidth]{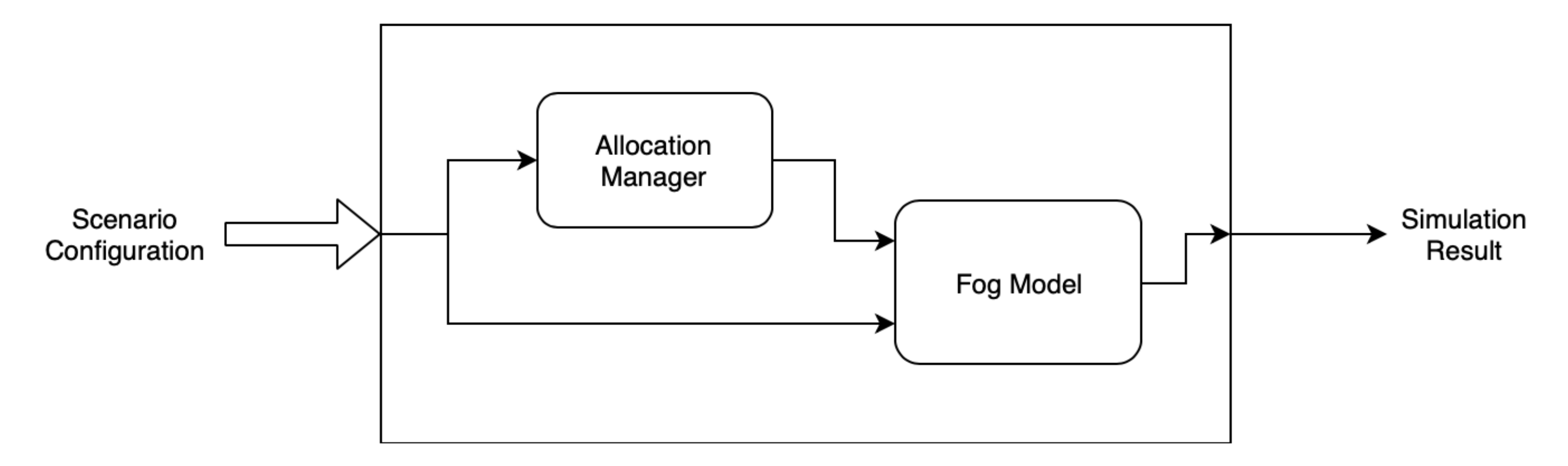}
        \caption{Mercury framework.}
        \label{fig:mercury}
\end{figure}

Mercury's fog model is flexible, configurable, and allows modeling complex and detailed scenarios with great detail. However, this tool may be used for straightforward, standard use cases. To make it easier for the user to configure the fog layers, Mercury offers an
Allocation Manager, a set of automatic tools to optimize the configuration of
the model to be explored. The allocation manager provides scenario
optimizations to find the most suitable locations for \glspl{AP} and
\glspl{EDC} depending on \gls{UE} modules' location histogram, as shown in
Figure~\ref{fig:mercury}.

The allocation manager splits the \gls{UE} location history into spatial cells
with length and width of a configurable spatial resolution. For each of these
cells, the allocation manager generates temporal snapshots using a given time
window. For each of these snapshots, the number of unique \gls{UE} nodes is
computed. Finally, the allocation manager generates a density map composed of
the snapshot with the highest number of \gls{UE} devices for each cell.  Then,
a modified K-Means algorithm~\cite{macqueen1967} is computed against the
resulting density map. The algorithm is tuned to find solutions with a similar
amount of \gls{UE} density per cluster, thus evenly dividing network traffic
among \glspl{AP}. Finally, the position of the \glspl{EDC} is obtained using
the standard K-Means algorithm on the \glspl{AP}' locations.

\section{Detailed Model Implementation}
\label{sec:4_implementation}
Mercury's model for simulating fog computing scenarios is built on top of \gls{DEVS}.
\gls{DEVS} is a general formalism for discrete event system modeling based on
set theory~\cite{Zeigler2000}. The \gls{DEVS} formalism provides the framework for
information modeling, which gives several advantages to analyze and design complex systems:
completeness, verifiability, extensibility, and maintainability. Once a system is described
in terms of the \gls{DEVS} theory, it can be easily implemented using an existing
computational library. There are two types of \gls{DEVS} models: atomic and coupled. Atomic
\gls{DEVS} models process input events based on their model's current state and condition.
They also generate output events and transition to the next state. Coupled models are
the aggregated composition of two or more atomic and coupled models connected by explicit
couplings.

\gls{DEVS} was selected since its formalism allows the specification of a modular and
hierarchical design, supports scalability and reusability, provides simple and clear
semantics for the basic model behavior, and represents an explicit separation between the
model specification and its corresponding implementation. To develop Mercury, we addressed
\gls{MBSE} methodologies. The elements and functionalities required for modeling the
scenario were identified, decomposed, and characterized before implementing them. 
In particular, system specification was supported with \gls{SysML} diagrams. \gls{SysML} is
a general-purpose graphical modeling language for analysis, specification, design,
verification, and validation of systems.
This section describes each of these implementations in detail. 

To implement the previously defined model, we built Mercury on top of
xDEVS~\cite{RiscoMartin2017}, a \gls{DEVS}-compliant package for C++, JAVA, and Python 3. In
this case, Mercury has been implemented using the xDEVS/Python API. Conceptually, \gls{DEVS}
separates models from the underlying simulator, making it possible to simulate the same model
using centralized, parallel, or distributed execution modes. In this regard, \gls{DEVS}
models are easily distributed since tasks can be executed in different nodes because these
tasks do not share memory. As a result, the net-centric xDEVS library allows a complex
\gls{DEVS} model to be executed in a sequential, parallel, or distributed manner, without
modifying the underlying Mercury model. The difference consists of running the simulation
using the same root \texttt{Coupled} model, but through the \texttt{Coordinator} class
(sequential simulation), the \texttt{CoordinatorParallel} class (parallel simulation using
threads), or the \texttt{CoordinatorDistributed} class (distributed simulation using
sockets).

\subsection{Structure and Behavior}

As stated in Section~\ref{sec:3_architecture}, Mercury's fog model is divided into six
different layers. Each layer contains submodules that define the layer's
behavior. This section explains the structure and behavior of the key submodules for each layer.

\subsubsection{Edge Federation Layer}
\label{subsubsec:edgefed}

\begin{figure}[htb]
	\center
	\includegraphics[width=\textwidth]{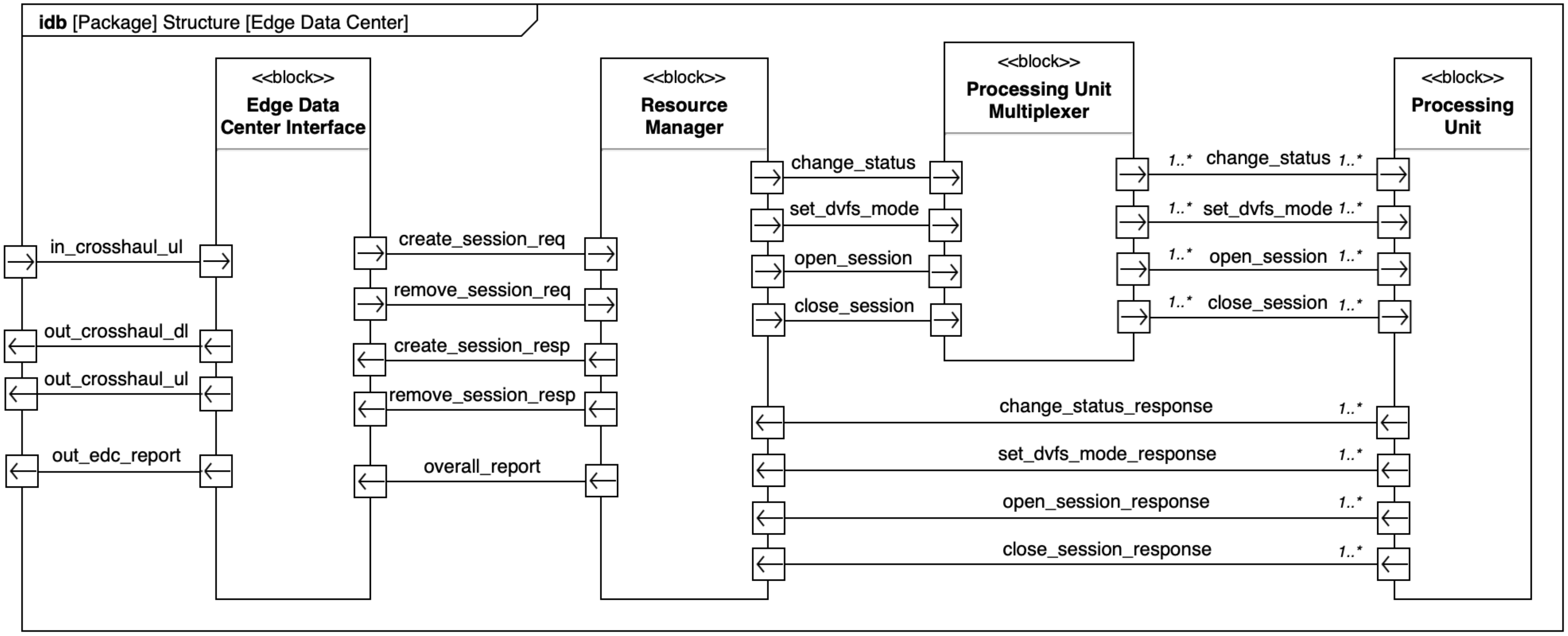}
	\caption{Edge Data Center internal block diagram.}
	\label{fig:edc_i}
\end{figure}

This layer contains all the \glspl{EDC} that compose the edge federation. The
edge federation controller monitors the status of all the \glspl{EDC} and reports it to
the \gls{ISP} for network management purposes. Every \gls{EDC} can be divided into three
different main submodules: processing unit(s), resource manager, and data center
interface. Figure~\ref{fig:edc_i} shows a \gls{SysML} \gls{IBD} of an
\gls{EDC}.

\begin{figure}[htb]
	\center
	\includegraphics[width=0.85\textwidth]{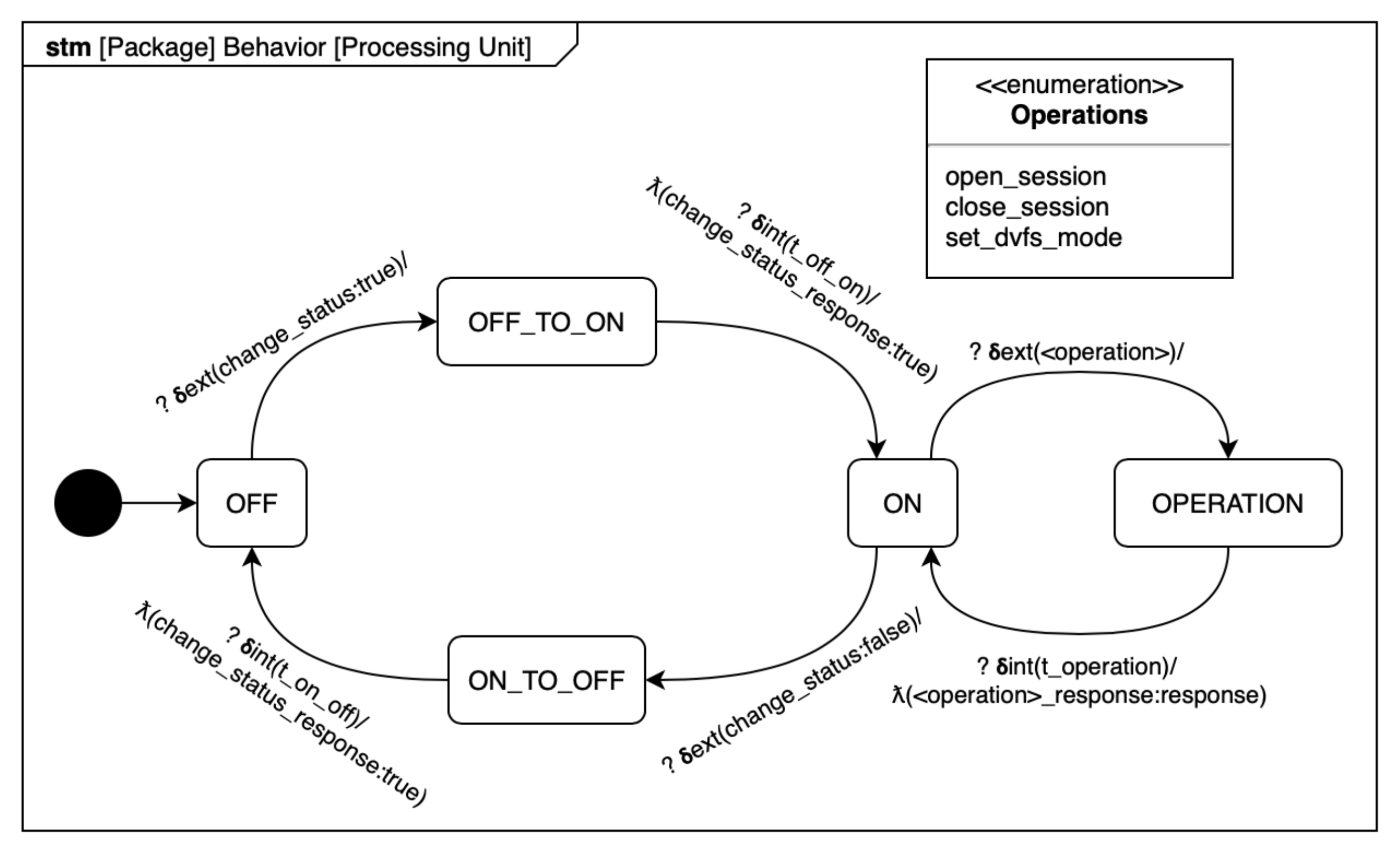}
	\caption{Processing unit state machine.}
	\label{fig:pu_stm}
\end{figure}

\paragraph{Processing Units}
These elements are hardware units - e.g., \glspl{GPU} - that provide computing resources to
an \gls{EDC}. A processing unit has a certain amount of computing resources depending on
its architecture - e.g., memory and working clock frequency. 
Also, it has a \gls{DVFS} table, which is a set of possible hardware configurations of the processing unit. Depending on
the selected hardware configuration, the processing unit's available resources and power
consumption may vary: setting energy-saving configurations diminishes the demanded power
for a processing unit to operate; however, its computing resources cannot be exploited
entirely, as these configurations - e.g., lower clock frequency or voltage - reduce the
number of instructions per second that can be performed. In general, depending on the
percentage of computing resources under use, a processing unit can change its \gls{DVFS}
configuration for reducing power consumption while meeting computing demand.
Figure~\ref{fig:pu_stm} shows the \gls{STM} that describes the behavior of a processing
unit.

The \gls{EDC} power consumption is modeled as the summation of the power consumed by all
the processing units in the \gls{EDC}. If a processing unit is powered off, its power
consumption is zero (see Equation~\ref{eq:puPoff}). However, it cannot host any session: a
processing unit is active only when switched on. Equation~\ref{eq:puPon} provides the
instantaneous power of a processing unit, where u(t) is its utilization - i.e.,
used computing resources at time t - and DVFS(t) is the \gls{DVFS} configuration used at
the same instant.

\begin{numcases}{P_{p.u.}(t) =}
	\label{eq:puPoff}
	0 & \text{if p.u. is powered off} \\
	\label{eq:puPon}
	\text{f(u(t), DVFS(t))} & \text{otherwise}
\end{numcases}


\paragraph{Resource Manager}
It is in charge of controlling all the processing units. When a session
creation request is received, the resource manager decides on which processing unit the
session is going to be hosted. 
It is possible to define different session dispatching algorithms, prioritizing power
consumption or \gls{QoS}, for instance. Mercury incorporates by default two different dispatching
algorithms: \texttt{MinimumWorkloadStrategy}, which forwards new sessions to the
processing unit with lower utilization, and \texttt{MaximumWorkloadStrategy}, which
dispatches new sessions to the processing unit with highest utilization - provided
that it has enough available computing resources for hosting the new session. However,
new dispatching algorithms can easily be
defined for tailored purposes. The resource manager is also responsible for managing the
\gls{DVFS} mode and switching the processing units on and off when required.

\paragraph{Data Center Interface}
This module controls the interaction with other submodules - i.e., \glspl{AP} and
\glspl{UE} -, adding networking capabilities to the \gls{EDC}. 
Messages from the \gls{EDC} are encapsulated as physical packets with
transmitted power, used bandwidth, spectral efficiency, and frequency carrier. The data
center interface also applies a transmission delay to messages depending on their data size.
On the other hand, physical packets received from the crosshaul layer are decapsulated, their
type is detected, and then they are forwarded through the correspondent port.

\subsubsection{Core Network Layer}
This layer gathers the tasks that are to be performed within the ISP's core network.
For accurately modeling a fog computing scenario, two functions were implemented: the
\gls{AMF}, which checks access permissions for any \gls{UE} that requests to connect to
the \gls{RAN}, and the \gls{SDN} controller, which configures the links between
\glspl{AP} and \glspl{EDC} based on proximity and \glspl{EDC}' availability.

\subsubsection{Access Points Layer}
\begin{figure}[b!]
	\center
	\includegraphics[width=\textwidth]{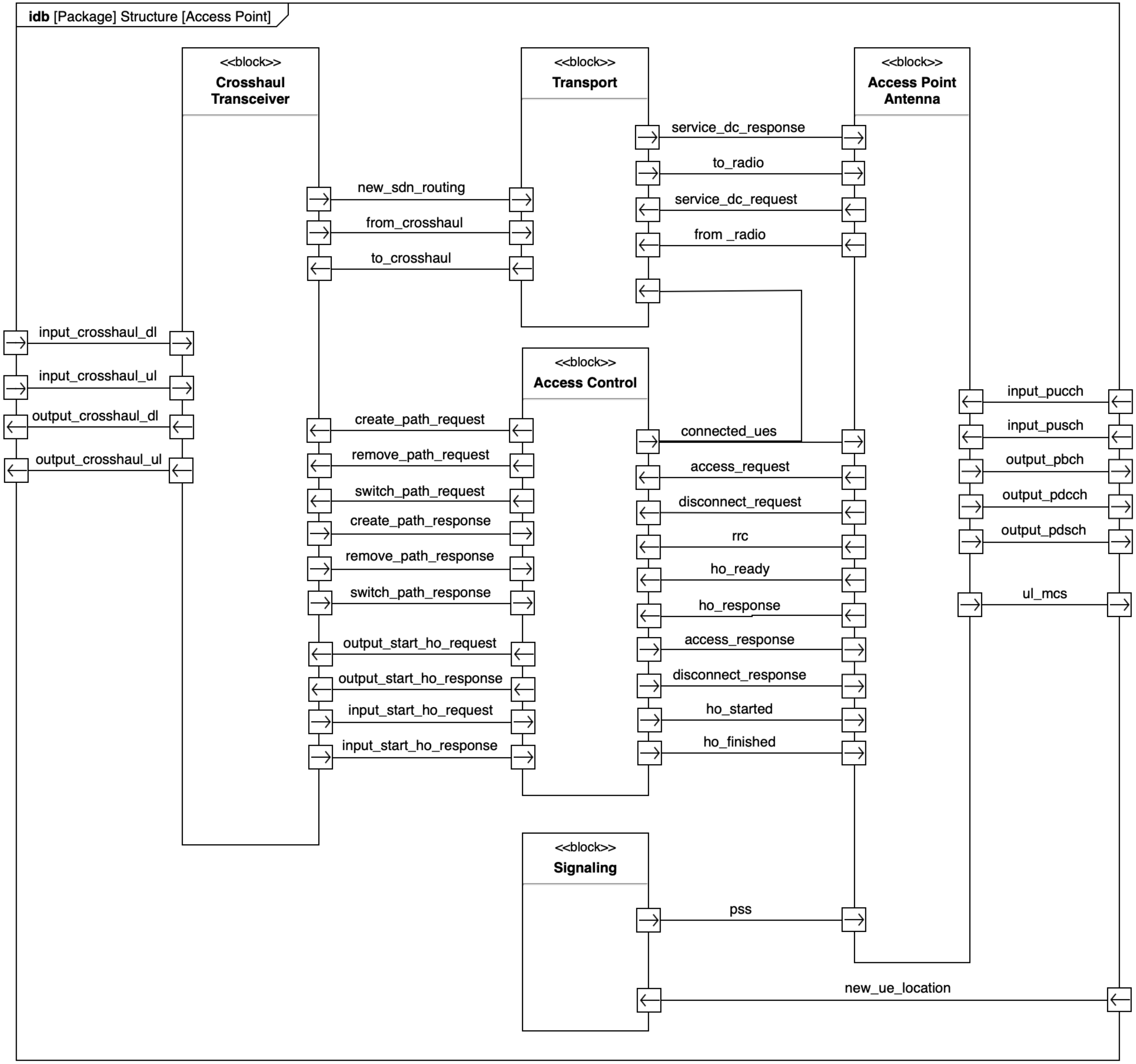}
	\caption{Access Point internal block description.}
	\label{fig:ap_i}
\end{figure}

This layer comprises all the \glspl{AP}.
Figure~\ref{fig:ap_i} shows the \gls{IBD} of an \gls{AP}.
The Signaling module is responsible for emitting \gls{PSS} messages for broadcasting the
\gls{AP}'s information. The Access Control module administers \gls{UE}'s connections and
triggers handover processes when required. Moreover, the Transport module handles the
routing paths of sessions between \glspl{UE} and \glspl{EDC}.
The Crosshaul Transceiver submodule controls the interactions with the \glspl{EDC} and the
core network functions. Its behavior is the same as the \gls{EDC}'s data center interface. 

\begin{figure}[htb]
	\center
	\includegraphics[width=0.8\textwidth]{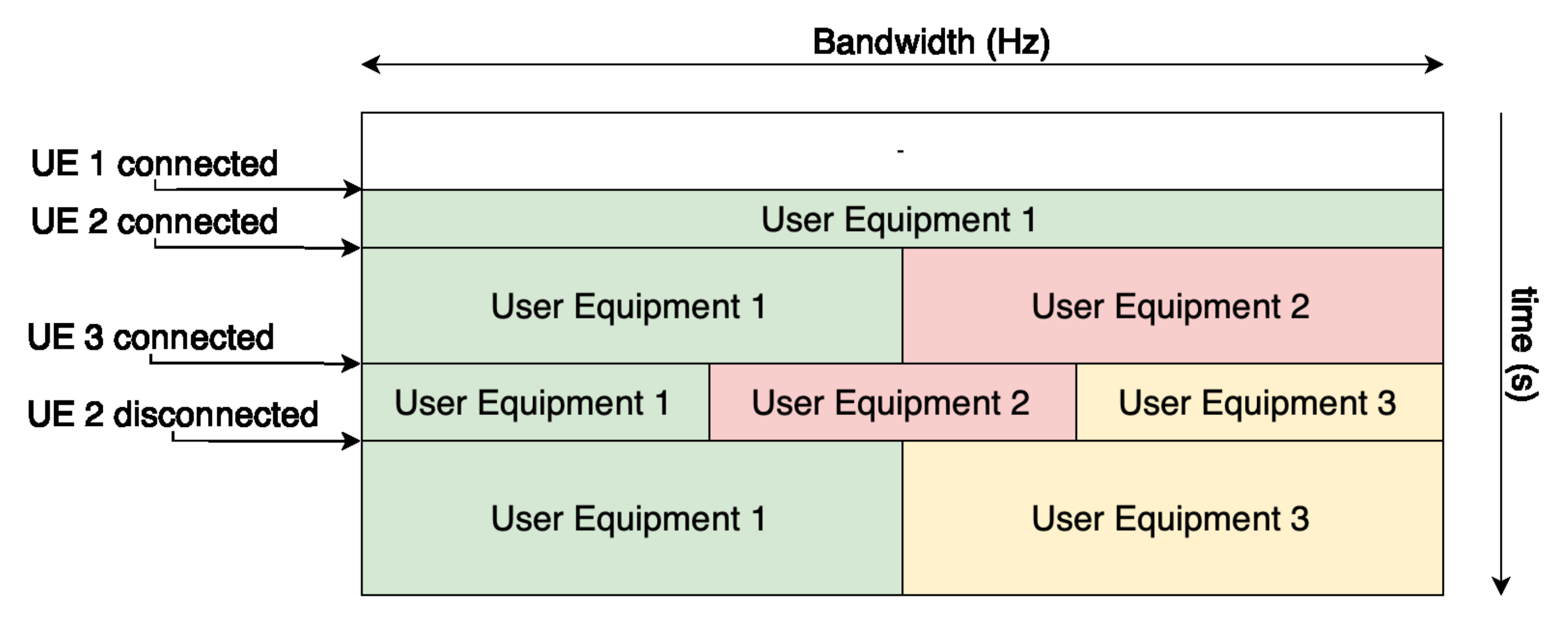}
	\caption{Access Point bandwidth assignment schematic.}
	\label{fig:ap_bw}
\end{figure}

Finally, the Access Point Antenna module adapts messages that interact with the radio
interface layer. Due to the particularities of radio interfaces, the behavior of this module
differs significantly from the crosshaul transceiver: first, the radio spectrum is shared
among every \gls{UE} connected to the \gls{AP}, and the bandwidth for each radio link may
vary: the more devices connected to an \gls{AP}, the lower available bandwidth for each of
them. Mercury incorporates a predefined bandwidth share algorithm that distributes the
available bandwidth evenly between the connected \gls{UE} nodes. However, it is possible to
define a custom bandwidth share algorithm. Figure~\ref{fig:ap_bw} shows a schematic of how
the predefined bandwidth share algorithm works.

Moreover, as \glspl{UE} may change their position, the radio link quality changes
during the simulation. This may lead to a variation in the \gls{MCS} used for
that channel, and therefore either improve or worsen the radio link's spectral
efficiency. The \gls{AP} antenna monitors the radio uplink quality of the \glspl{UE}
connected to the \gls{AP} and determines which \gls{MCS} must be used depending on the
perceived \gls{SNR}. The available \glspl{MCS} can be defined by the user. However, by
default, the \gls{NR} \gls{MCS} table is used~\cite{3gpp38214}.

\subsubsection{IoT Devices Layer}

\begin{figure}[b!]
	\center
	\includegraphics[width=\textwidth]{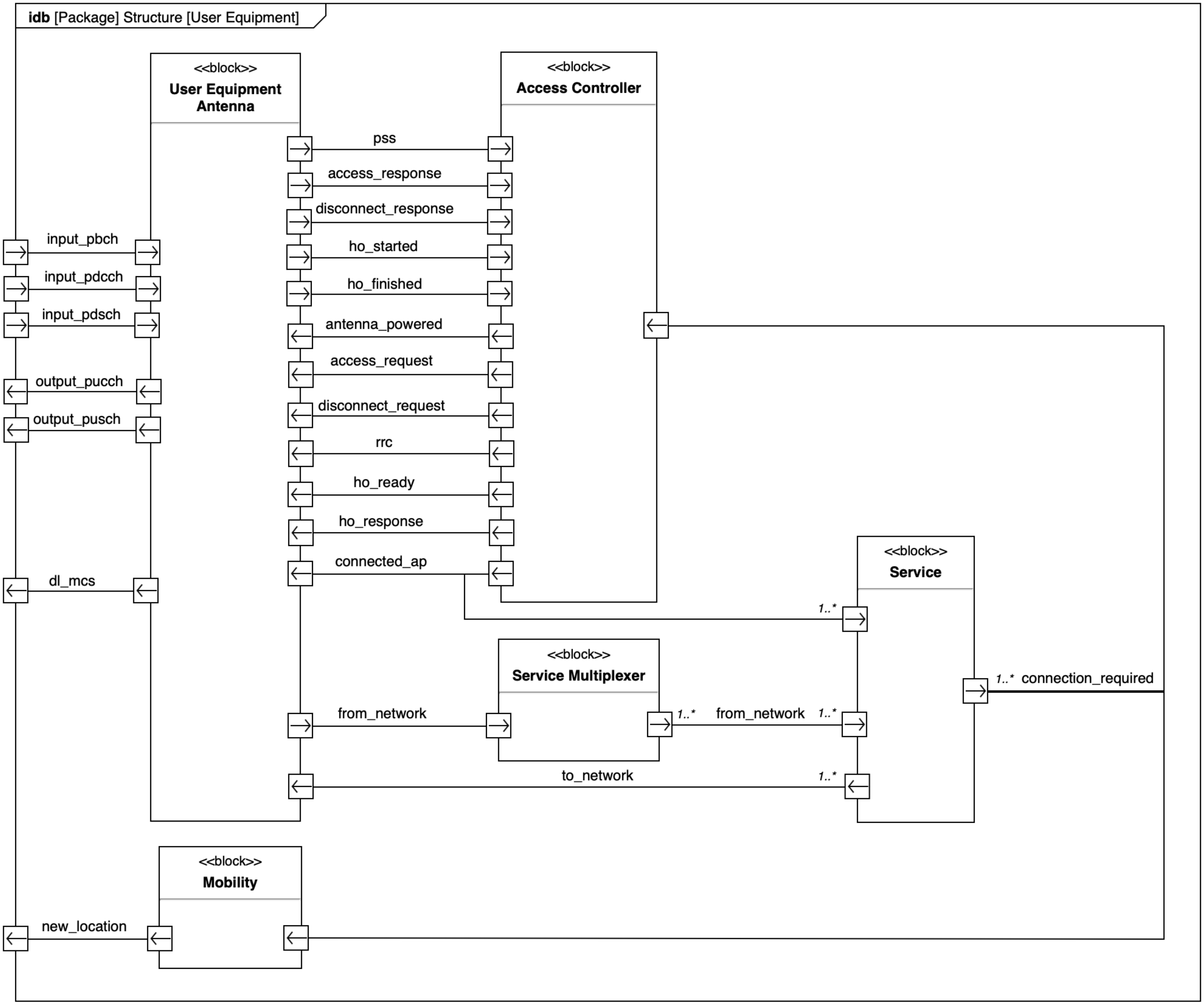}
	\caption{User Equipment internal block description.}
	\label{fig:ue_i}
\end{figure}

\begin{figure}[htb!]
	\begin{subfigure}{\textwidth}
		\center
		\includegraphics[width=\textwidth]{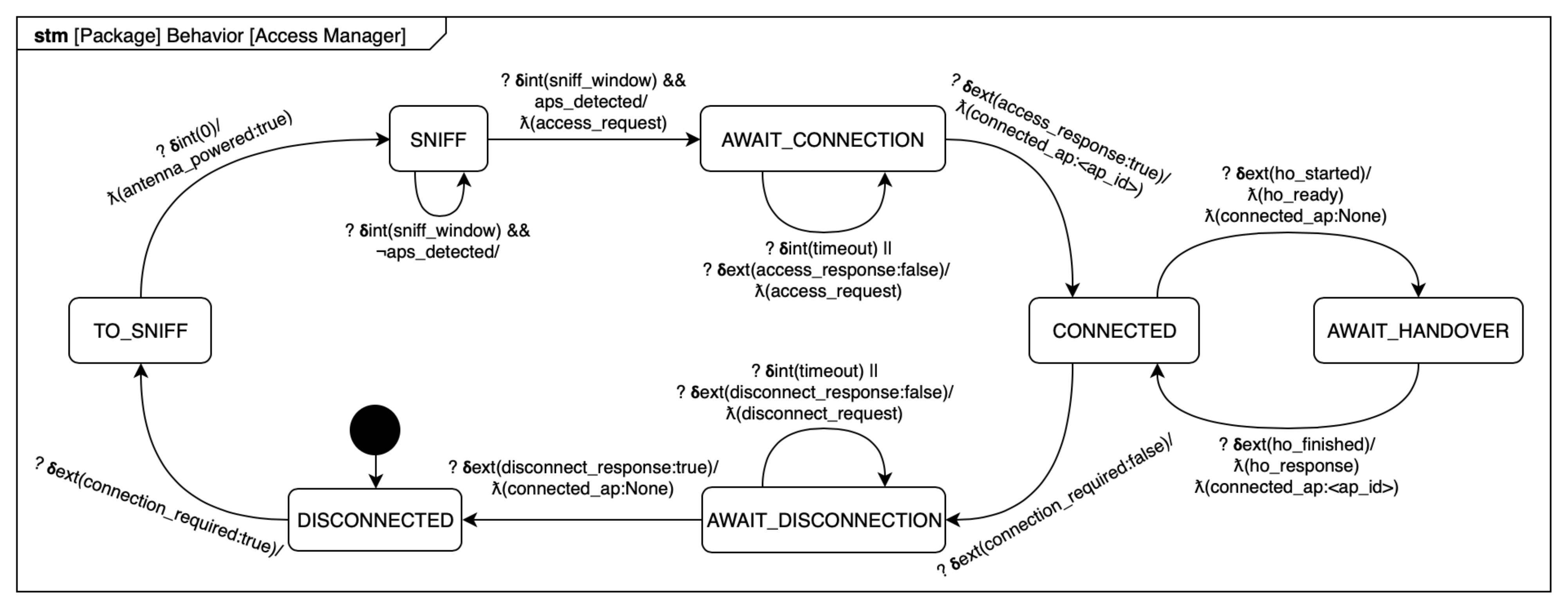}
		\caption{UE access controller state machine.}
		\label{fig:access_stm}
	\end{subfigure}
	\begin{subfigure}{\textwidth}
		\center
		\includegraphics[width=\textwidth]{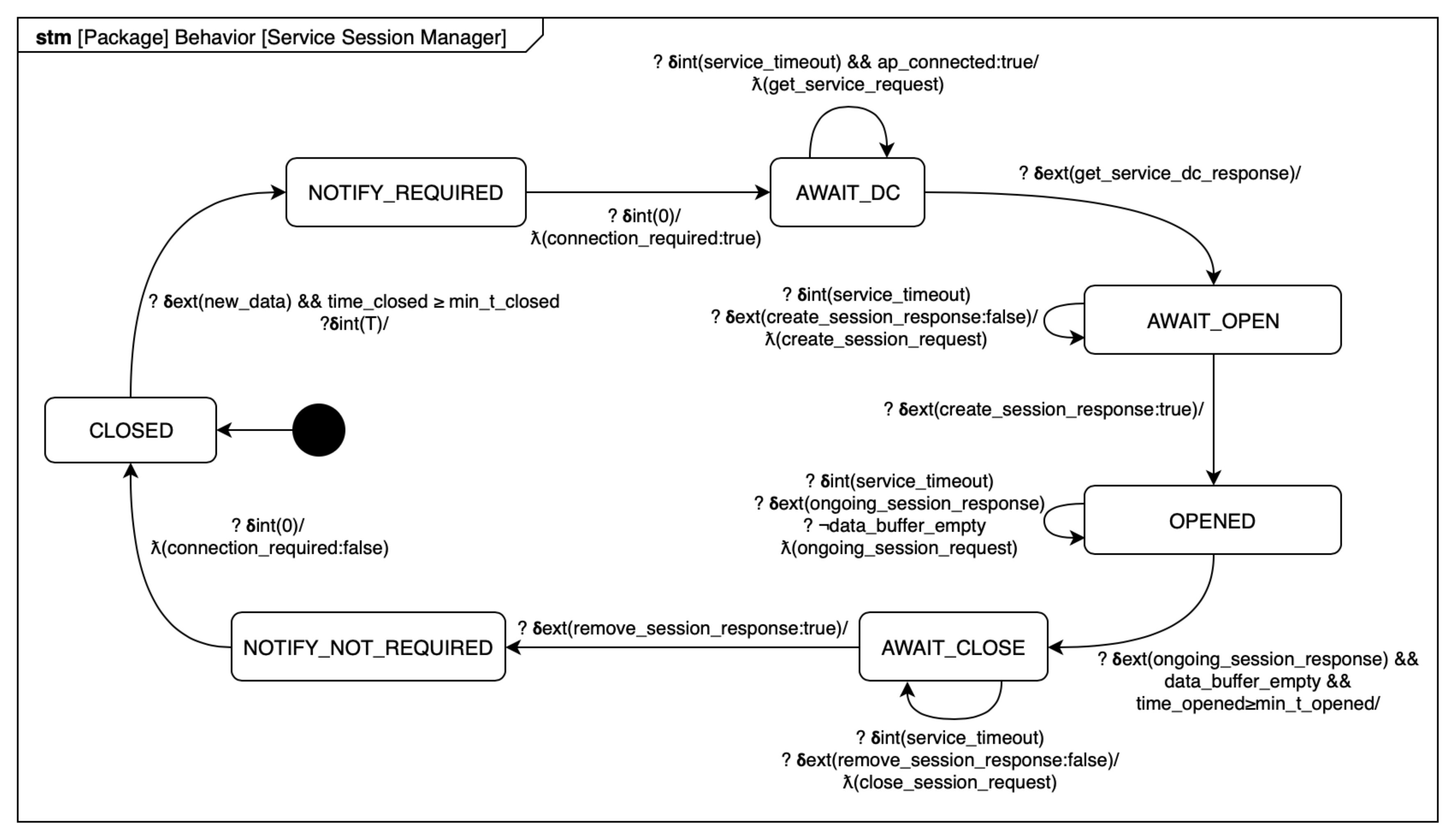}
		\caption{UE service state machine.}
		\label{fig:service_stm}
	\end{subfigure}
	\caption{User Equipment-related state machines.}
\end{figure}

This layer contains the \glspl{UE} that are defined in the scenario. Each
\gls{UE} is divided into five submodules, as shown in Figure~\ref{fig:ue_i}.
The mobility module is in charge of managing the \gls{UE}'s location. The \gls{UE} antenna
is similar to the \gls{AP}'s, but it monitors the downlink quality instead of the uplink,
and it does not configure the available bandwidth but adapts to decisions made by the
\glspl{AP}. The access controller is responsible for connecting to the
\gls{RAN}. It discovers available \glspl{AP}, starts the access sequence, and handles
handovers. Figure~\ref{fig:access_stm} shows the \gls{STM} that models the behavior of the
access controller. Finally, each \gls{UE} has one or more service submodules. Each service
submodule models the behavior of an \gls{IoT} application service.
Figure~\ref{fig:service_stm} shows the \gls{STM} that models the behavior of a service.

\subsubsection{Crosshaul Layer}
This layer models a 5G-based transport network that combines backhaul and fronthaul,
enabling a flexible software-defined reconfiguration of all the networking
elements~\cite{crosshaul}. The crosshaul layer allows event transmission between
Access Points, Core Network, and Edge Federation layers. The objective of this layer is
twofold: on the one hand, it applies a power attenuation function to all the messages that
go through it. Besides, it is also in charge of applying the corresponding propagation
delay to communications. The crosshaul layer is composed of two channels that
conform a \gls{FDD} full-duplex channel for data transmission between modules: one for
uplink and other for downlink communications.

\subsubsection{Radio Interface Layer}
This layer interconnects the Access Points and \gls{IoT} devices layers. Its behavior is
similar to the crosshaul layer. We considered the \gls{NR} physical
interface~\cite{nrqualcomm} as a base for modeling this layer. Five physical radio channels
compose the radio interface: one of them is the \gls{PBCH}, which is used by \glspl{AP} for
broadcasting \gls{PSS} messages. The \gls{PUCCH} and \gls{PDCCH} are used for transmitting
control messages - e.g., access requests and responses or handover-related messages.
Finally, the \gls{PUSCH} and \gls{PDSCH} constitute a \gls{FDD} full-duplex channel for data
transmission between \glspl{AP} and \gls{UE}. 

\subsection{Inter-module Communications}

In this subsection, the interactions between the main submodules of Mercury's fog model are explained.

\subsubsection{Radio Access Network Connectivity}

\begin{figure}[p!]
	\begin{subfigure}{\textwidth}
	\center
	\includegraphics[width=0.88\textwidth,height=0.33\textheight]{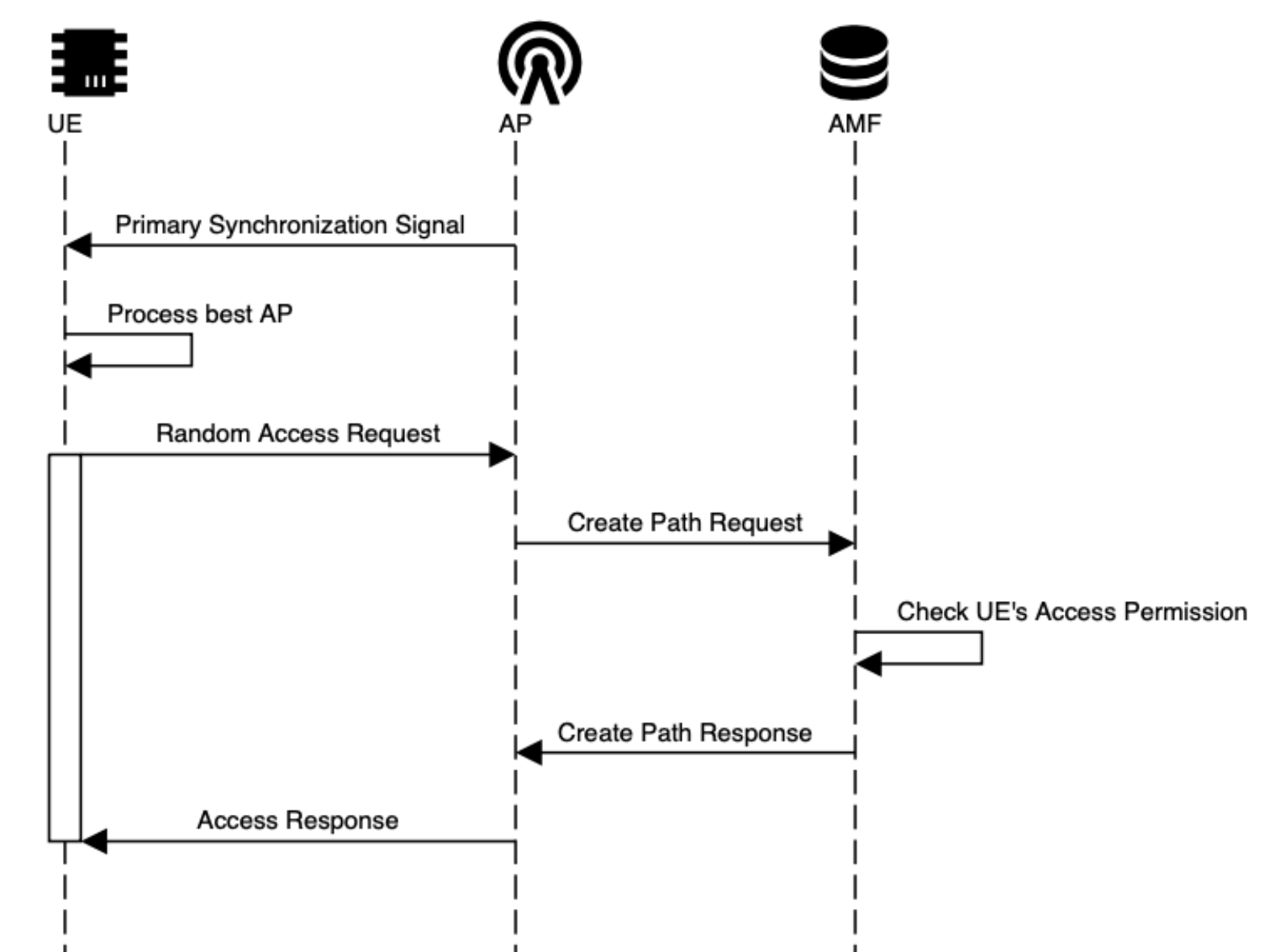}
	\caption{UE access request sequence diagram.}
	\label{fig:ue_connect}
	\end{subfigure}
	\begin{subfigure}{\textwidth}
		\center
		\includegraphics[width=\textwidth,height=0.55\textheight]{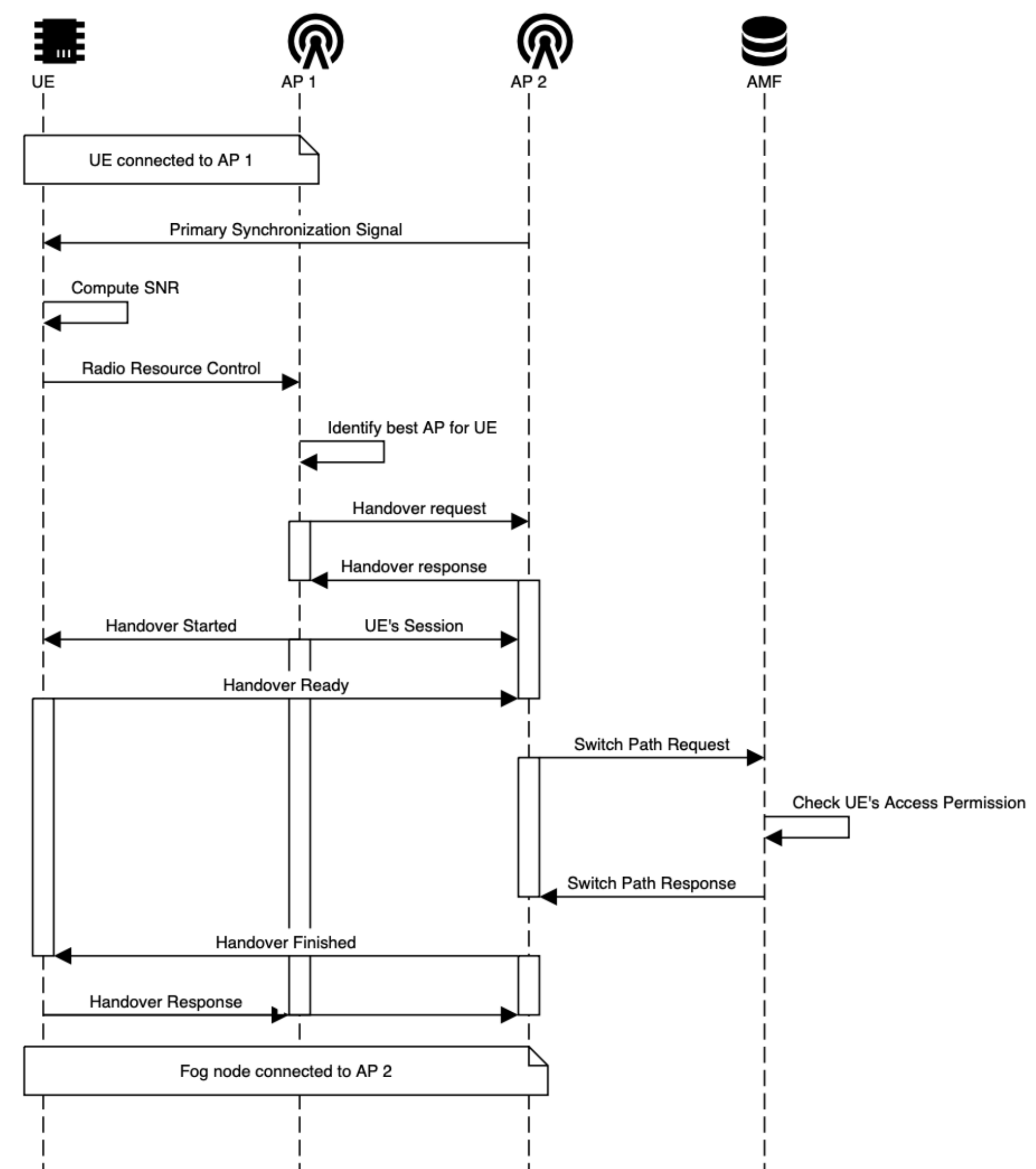}
		\caption{UE handover sequence diagram.}
		\label{fig:handover}
	\end{subfigure}
\caption{UE connection sequence diagrams.}
\label{fig:ue_connects}
\end{figure}

\gls{UE} need to be connected to the \gls{RAN} to successfully establish a service session
with an \gls{EDC} of the edge federation. \gls{UE} connect to the \gls{RAN} via the
\glspl{AP} that are scattered around the scenario. First, every \gls{UE} listens to
\gls{PSS} messages from all the \glspl{AP}. These messages contain information about the
emitter \gls{AP} and enable the \gls{UE} to detect which \gls{AP} offers the best radio
signal quality. Eventually, \gls{UE} send an access request message to their most suitable
\gls{AP}. \glspl{AP} forward these requests to the \gls{ISP}'s core network, where the
\gls{AMF} checks the access control policies before granting the access. Once the \gls{UE}
receives an affirmative response to its access request, it is connected to the \gls{RAN} via
the selected \gls{AP}. Figure~\ref{fig:ue_connect} shows a sequence diagram of the
connection process. For disconnecting from the \gls{RAN}, the communication sequence is
similar.

As \gls{UE}'s location may change over time, the most suitable \gls{AP} can also change.
Hence, \gls{UE} modules send periodic \gls{RRC} messages to their correspondent \gls{AP}. If
the \gls{AP} finds another more suitable \gls{AP}, it triggers a handover process for
transferring the \gls{UE}'s session to the new \gls{AP}. Figure~\ref{fig:handover} presents
the sequence diagram for the \glspl{AP} handover process.

\subsubsection{Service-Related Communications}

Once the connection to the \gls{RAN} has been established, the \gls{UE} proceeds to ask to
its \gls{AP} for the most suitable \gls{EDC} for each of the implemented applications.
\glspl{AP} have an internal list of the most suitable \gls{EDC} depending on the application
requested. This table is generated by the \gls{SDN} controller, which is in charge of
providing network slicing capabilities to the \gls{RAN}. The table changes dynamically
depending on the federated \glspl{EDC}' available resources.

\begin{figure}[h!]
	\center
	\includegraphics[width=0.93\textwidth,height=0.8\textheight]{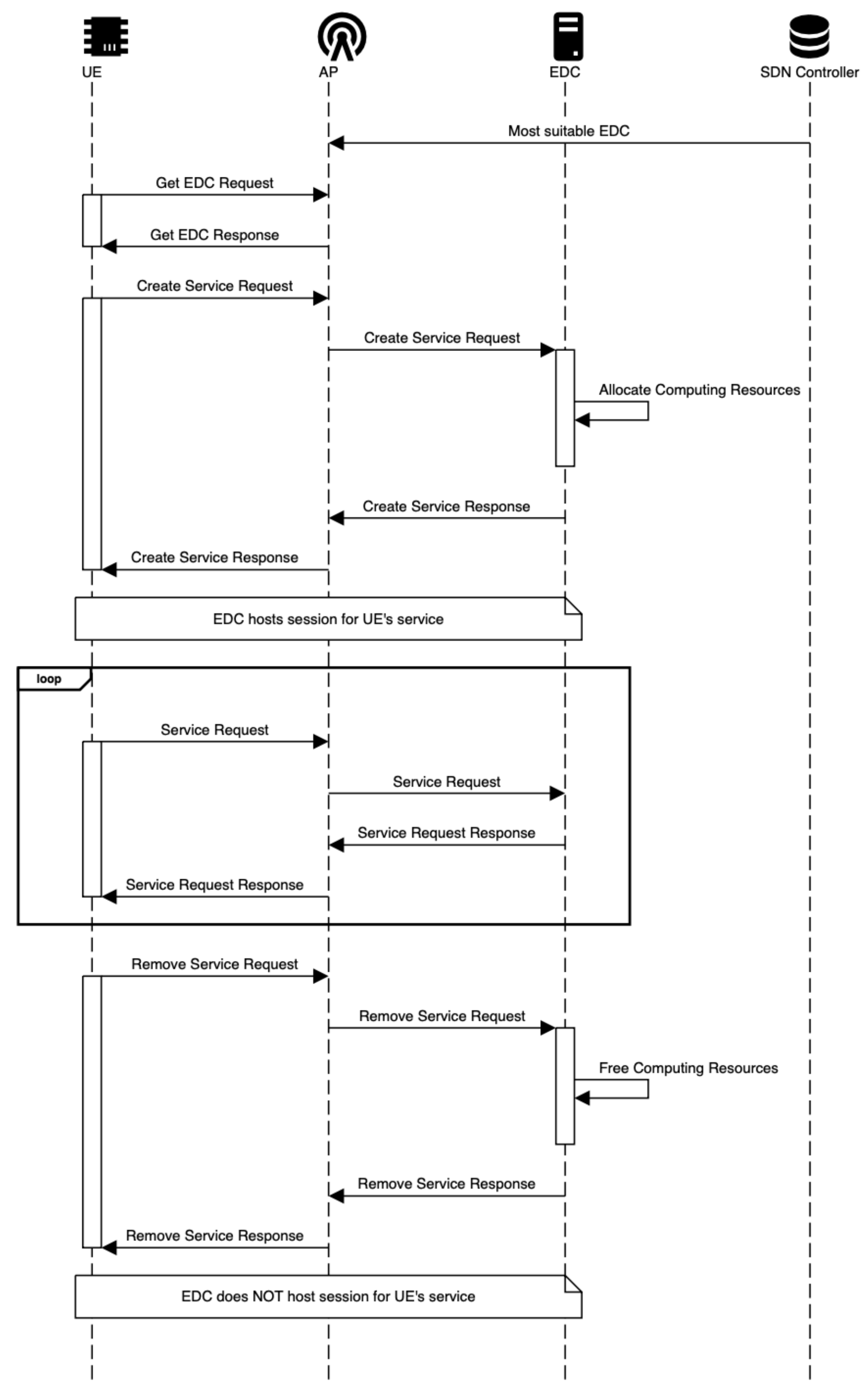}
	\caption{Service operations sequence diagram.}
	\label{fig:service_seq_diagram}
\end{figure}

\gls{UE} then send one session creation request per service to the
correspondent \gls{EDC}.
The \gls{EDC} that receives the request checks if there are enough resources for the new
service session request, and allocates the resources if available. The \gls{EDC} sends an
affirmative response to the \gls{UE} so it can transfer the service-related data stream,
which is processed in real-time. The \gls{UE} voluntarily closes the session after a
timeout to avoid resource starvation, repeating the entire service request process. Note
that the highest delays experienced by the \gls{UE} occur when creating and removing a
session. Once a session is opened, requests are quickly processed.
Figure~\ref{fig:service_seq_diagram} shows a sequence diagram of a single service.

\section{Use Case}
\label{sec:5_use_case}
In this section, we propose a use case scenario to show how Mercury can assist with the
dimensioning and operation tasks during the deployment of new edge computing
infrastructures. The output generated by the \gls{MS} framework is analyzed to give a sense
of how it can be interpreted to help on the decision-making process. Finally, we show how
the proposed scenario could be explored using other \gls{MS} frameworks to stress the
novelty and major advantages of Mercury.

\subsection{Scenario Description}

\begin{figure}[htb]
	\center
	\includegraphics[width=\textwidth]{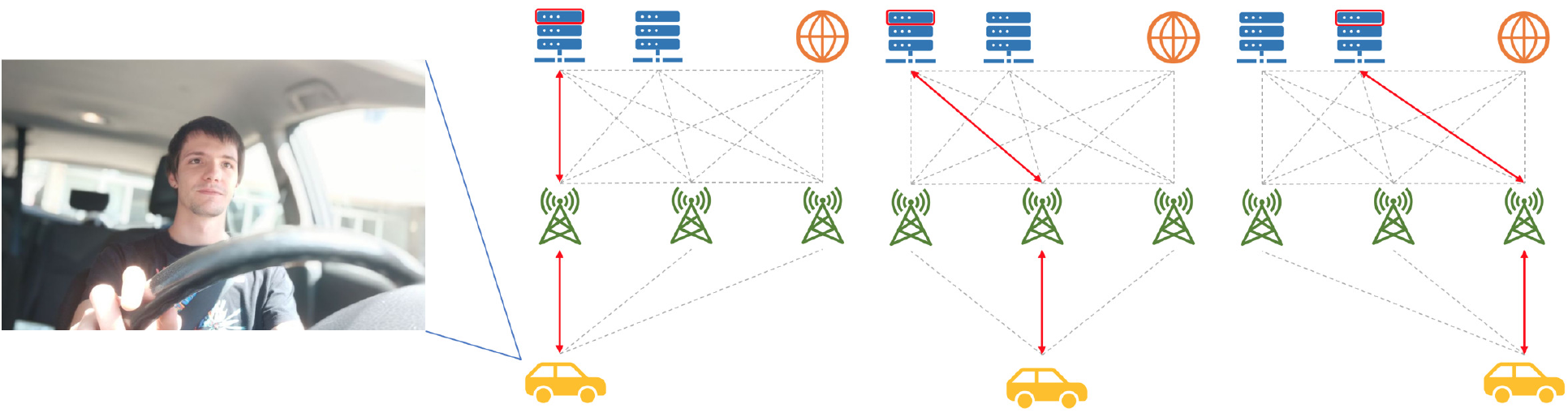}
	\caption{Use case schematic.}
	\label{fig:usecase}
\end{figure}

To analyze the performance of the model presented in this research, we propose a driving
assistance scenario. \gls{ADAS} scenarios provide a dynamic real-time data stream analytics
environment that handles large volumes of data. Each vehicle is running an onboard
predictive model that detects potentially dangerous situations according to real-time
images from the driver's face. The purpose of this use case is to
keep training the predictive model for each vehicle to capture the particularities of every
driver, improving the accuracy of the predictive model in a personalized way. Training
machine learning models is a computationally expensive process, and using computation
offloading could improve the overall performance while reducing the service deployment
cost: pools of specialized hardware resources are shared between all the users of the
application, thus making unnecessary to include extra expensive hardware in the vehicle.
The usage of an edge federation for computation offloading benefits from reducing latency
and core network traffic, as data is sent to \glspl{EDC} that are closer to the data
sources. Figure~\ref{fig:usecase} shows a schematic of the proposed use case.
\glspl{UE} (vehicles) periodically send images in real-time to be processed by \glspl{EDC},
which
train a custom predictive model for each vehicle, thus reducing the risk of an accident
occurring. To do so, \glspl{EDC} inject in real-time the new images sent by vehicles to
the training process.
Once the trained predictive models are significantly improved compared to the onboard versions, the embarked models can be upgraded during run-time. Note that each
vehicle is in motion and that the model will manage the necessary \gls{AP} and \gls{EDC}
allocation changes to ensure \gls{QoS} in the image data flow.

\begin{figure}[htb]
	\center
	\includegraphics[width=0.8\textwidth]{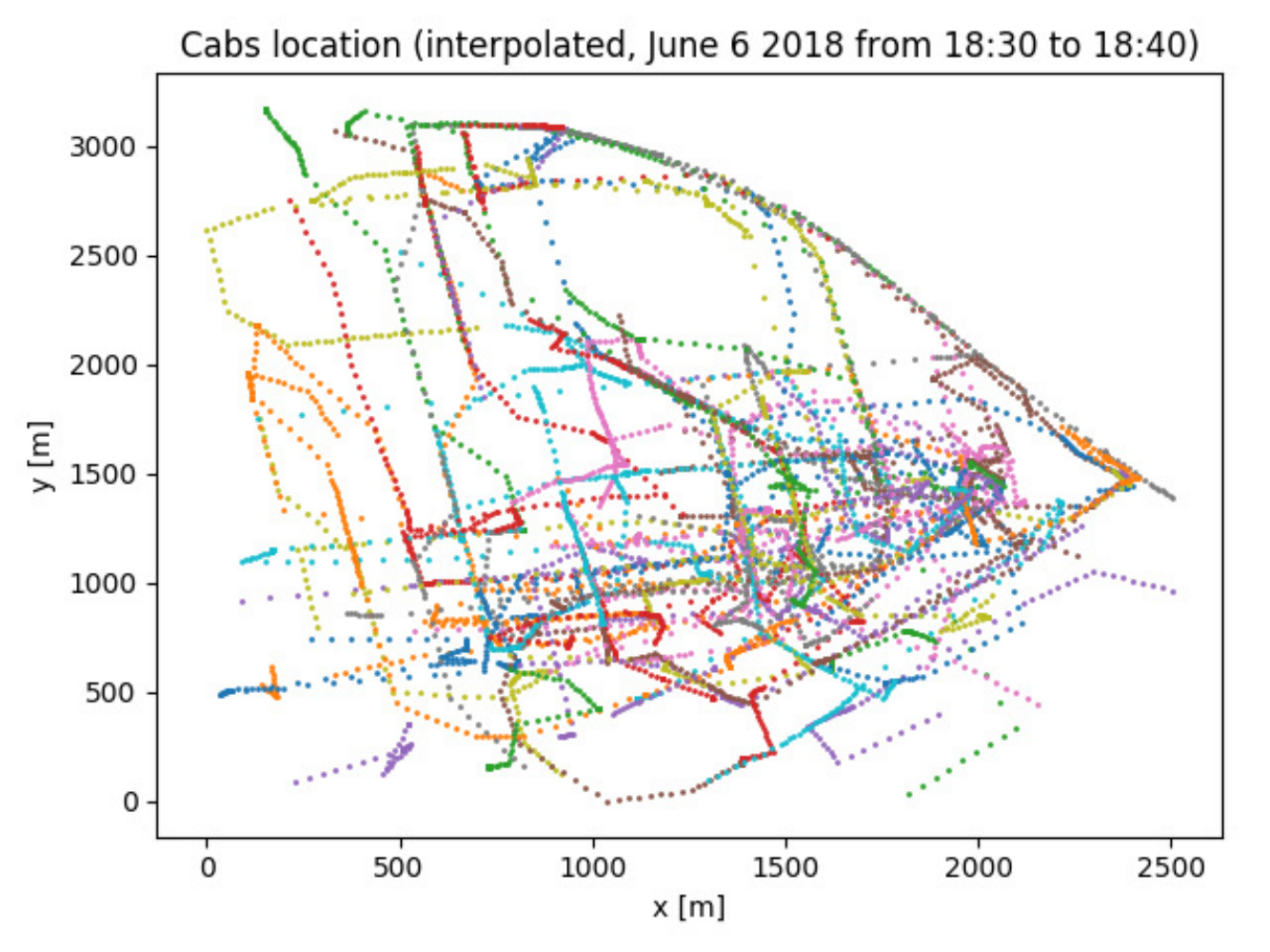}
	\caption{Taxis location during simulation time.}
	\label{fig:friscocabs}
\end{figure}

To present a realistic scenario, we feed our model with real mobility traces
provided by EPFL (Piorkowski et al.~\cite{comsnets09piorkowski}). The dataset
contains mobility traces of taxis in San Francisco, USA. The trace set provides 30
days information about \gls{GPS} coordinates of 500 taxis. Due to the high volume of
data, we constrain the dataset to the area of the bay during the busiest 10 minutes
of the peak hour of the day with the highest traffic flow (18:30:00 to 18:40:00 on 6
June 2008, PDT, UTC -7:00, daylight saving time). Our  resulting dataset presents 100
vehicles that are in our target area during all the simulation time. As the original
data are presented at a rate of approximately one trace per minute, we have performed
a linear interpolation to obtain a final rate of one trace approximately every 10
seconds to increase the granularity of the scenario. Finally, we transform \gls{GPS}
data to cartesian coordinates with an error lower than 10 centimeters per kilometer.
Figure~\ref{fig:friscocabs} shows the taxis' interpolated location history during the 10
minutes to be simulated in cartesian coordinates.

Regarding the \gls{ADAS} service, each vehicle sends a large volume of data
corresponding to a real workload from the Elektra Autonomous Vehicle
database~\cite{DiazChito201698}. We use the CVC11 Driver Face dataset,
which compiles Standard Definition (SD) images from onboard cameras using a
resolution of 640x480 pixels. The images feature male and female driver's
faces while driving in real scenarios. These images are processed in the \glspl{EDC}
computing infrastructure to train a classification model for accident prevention.
The required binary rate is modeled as 1 Mbps.

The \gls{ADAS} service configuration parameters are as follows: after idling during one second, each \gls{UE} sends a ``create service"
	request. If the \gls{UE} does not get a response after 0.35 seconds, it resends
	the request. After creating the service, each \gls{UE} sends a ``service request"
	message every second. These messages have 1 Mbit size, thus
	resulting in a binary rate of 1 Mbps per vehicle. Once 20 messages (around 20 seconds) are transmitted and replied,
	\gls{UE} nodes demand a session removal and idle for another second. Each \gls{UE}
	repeats the entire process until the end of the simulation.

The \glspl{EDC}' computing resources consist of a set of \glspl{GPU} of the Sapphire
Pulse Radeon RX 580 series. \gls{GPU}-based clusters are one of the best options to
provide the workload requirements demanded by \gls{ADAS} applications, as they run
\glspl{DNN} and \glspl{CNN}, frequently used for training classification models, with a
significantly higher performance than \glspl{CPU}~\cite{8057318}. The workload
consists of training the classification model, including the received images, which is
based on a CIFAR-10  model\footnote{https:/keras.io/examples/cifar10\_cnn/}. The
images used for training the model are from the \gls{ADAS} dataset from Elektra
Autonomous Vehicle\footnote{adas.cvc.uab.es/elektra}. Each \gls{GPU} can host the
sessions of up to five vehicles simultaneously - i.e., sessions require 20\% of
its computing resources. 

We use a \gls{GPU} power model based on an \gls{ANN} that depends on the current \gls{ADAS} workload.
The performance of this power model presents a
\gls{NRMSD} of 2.45\% and an $R^2$ of 99.01\%. The time required for powering on/off
the \glspl{GPU} is set to 1 second. For starting or removing a session, a \gls{GPU}
needs 0.2 seconds. 


\glspl{AP} antennas transmitting power was set to 50 dBm, their
gain to 0 dB and their equivalent noise temperature to 300 K. \gls{UE}'s gain and
equivalent noise temperature coincided with the \glspl{AP}', but their transmitting
power was limited to 30 dBm. Regarding the radio interface frequency band, the default value provided by the allocation manager is used, which is the n77 band - carrier
frequency of 33,000 MHz and bandwidth of 100 MHz per channel.

\subsection{Scenario Optimization}
\begin{figure}[htb!]
  \center
  \includegraphics[width=0.8\textwidth]{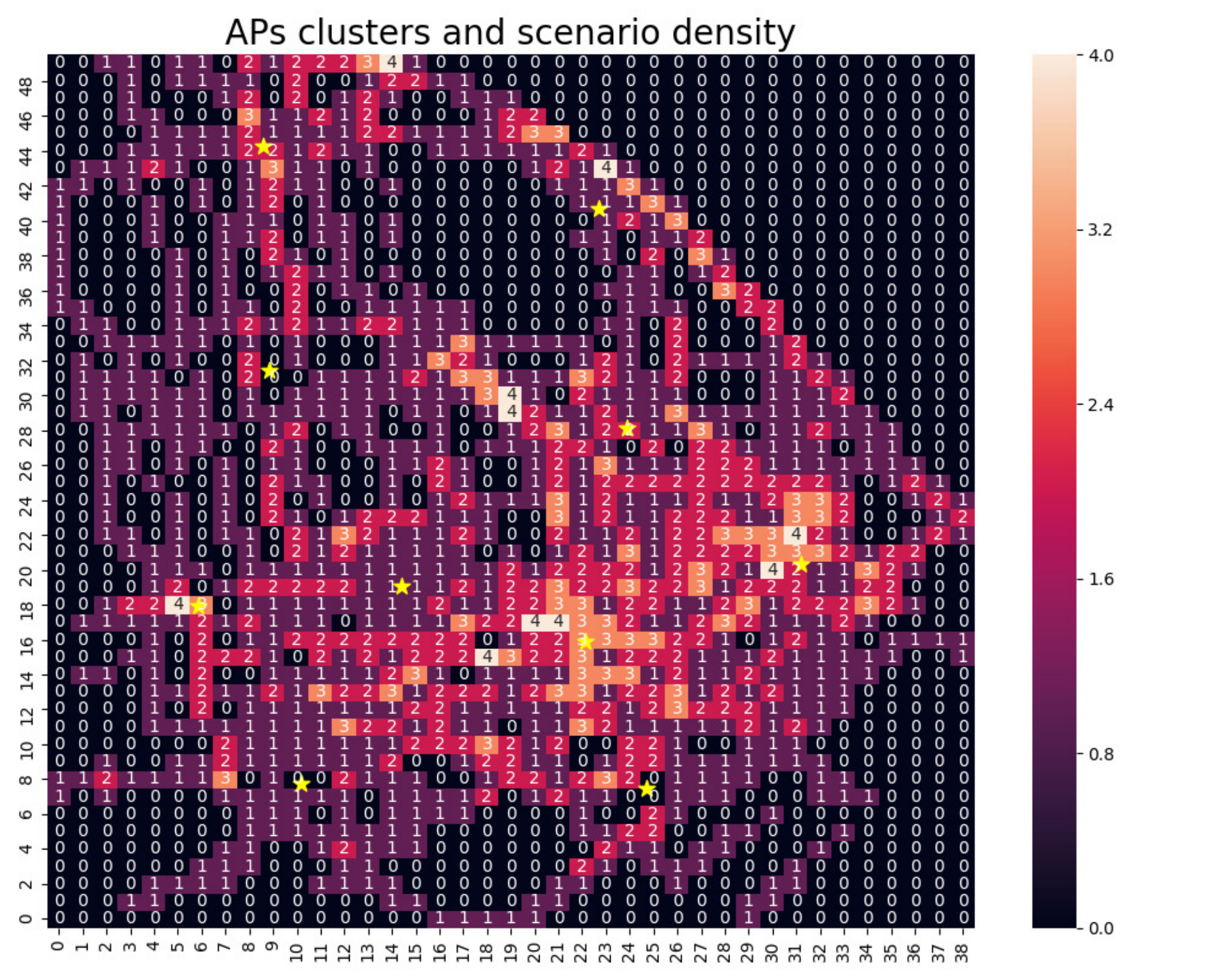}
  \caption{San Francisco scenario density.}
  \label{fig:cabdensity}
\end{figure}

Due to the scenario's complexity, we used Mercury's allocation manager
automatic tools. The number of \glspl{AP}, as well as their location, were
automatically set by the allocation manager. With a time window of 1 second and a
spatial resolution of 40 meters, the allocation manager built the density map shown in
Figure~\ref{fig:cabdensity}, placing 10 \glspl{AP} in the locations spotted with
yellow stars.

\begin{figure}[htb]
	\center
	\includegraphics[width=0.8\textwidth]{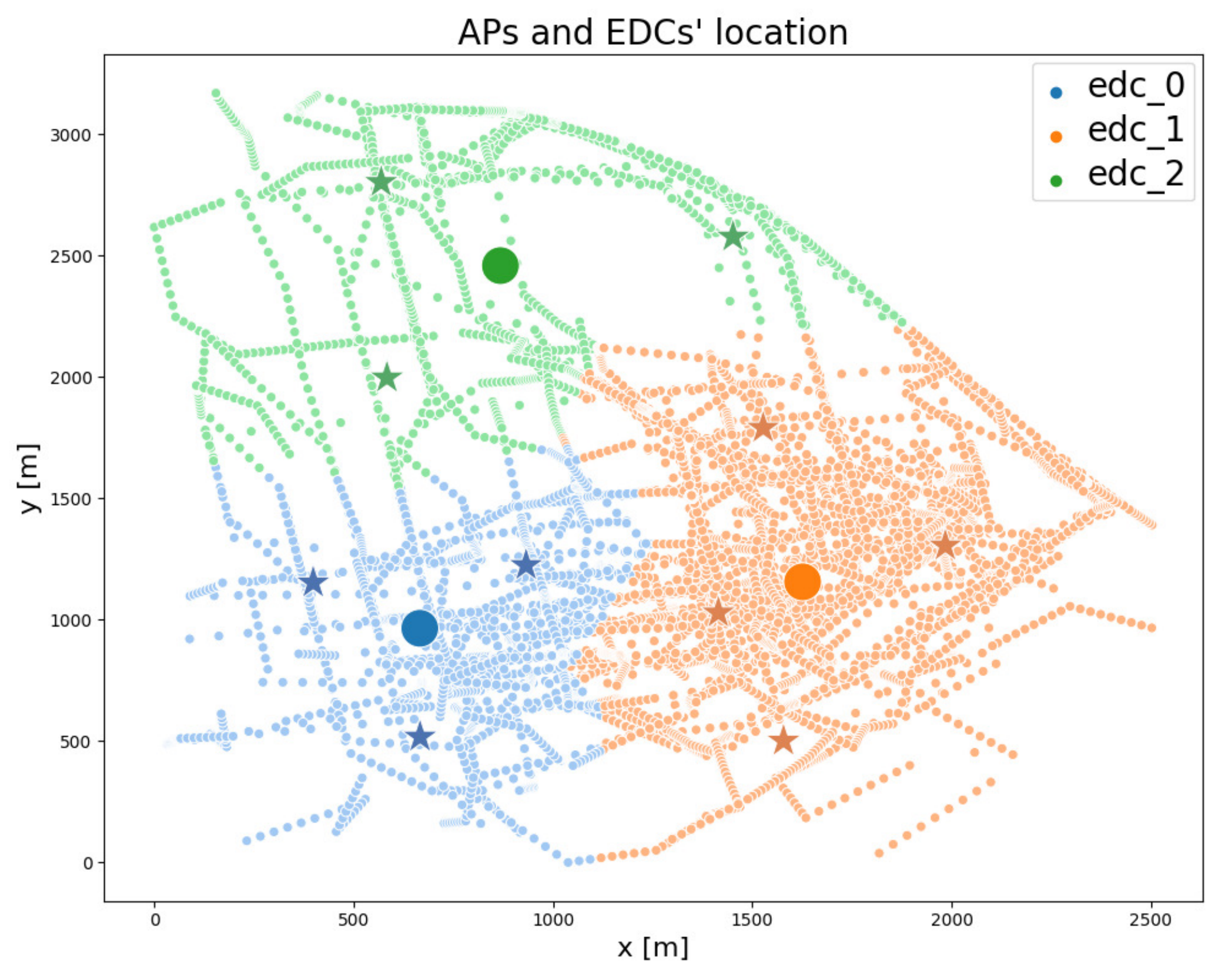}
	\caption{San Francisco final scenario.}
	\label{fig:friscoscenario}
\end{figure}

Regarding the \glspl{EDC}, the allocation manager determined that each \gls{EDC}
should have 20 \glspl{GPU} (5 sessions/GPU $\times$ 20 GPUs = 100 sessions). As
we set the replication factor to 3 - value for ensuring a high availability
infrastructure -, the scenario ended up with three \glspl{EDC}.
The final scenario is shown in Figure~\ref{fig:friscoscenario}, where thick dots
present the \glspl{EDC}, and the \glspl{AP} are displayed as stars.
The colors of the dots and the stars define the \glspl{EDC}' preferred activity
region.

\subsection{Experiments Definition}

To show how Mercury can assist on the analysis of the dynamic operation strategies for edge federations, we present
two different experiments that differ in the
\glspl{EDC}' resource manager configuration. The target of the first experiment is to
reduce the delay perceived by \gls{UE}. For doing so, all the \glspl{GPU} were powered
on during the whole simulation. The selected dispatching algorithm was the
\texttt{MinimumWorkloadStrategy}, as it was the one that reported the best \gls{QoS}
results in previous use cases~\cite{acceptedSummersim}. On the other hand, the second
experiment keeps switched off idle \glspl{GPU} and uses the
\texttt{MaximumWorkloadStrategy} for reducing overall power consumption. Both dispatching strategies are presented in Section~\ref{subsubsec:edgefed}.
Table~\ref{tab:frisco_conf} resumes the selected configurations for both experiments. Simulations were executed using PyCharm Professional 2019.1.1~\cite{pycharm} on a
MacBook Pro (Retina, 15-inch, Mid 2015) with a 2.5 GHz Intel Core i7 processor and a
16 GB 1600 MHz DDR3 memory. Each simulation took approximately 10 hours and 30
minutes.
   
\begin{table} [htb]
	\footnotesize
	\centering
	\caption {Configuration of San Francisco experiments.}
  	\label{tab:frisco_conf}
	\begin{tabular}{lccc}
  		\toprule
						& Unused Hardware Strategy & Dispatching Strategy 	\\
  		\midrule
		Experiment I	& Powered on 	& \texttt{MinimumWorkloadStrategy}  \\
		Experiment II	& Powered off 	& \texttt{MaximumWorkloadStrategy} 	\\
  		\bottomrule
  	\end{tabular}
\end{table}

\subsection{Simulation Results}
\begin{figure}[h!]
  \center
  \includegraphics[width=0.65\textwidth]{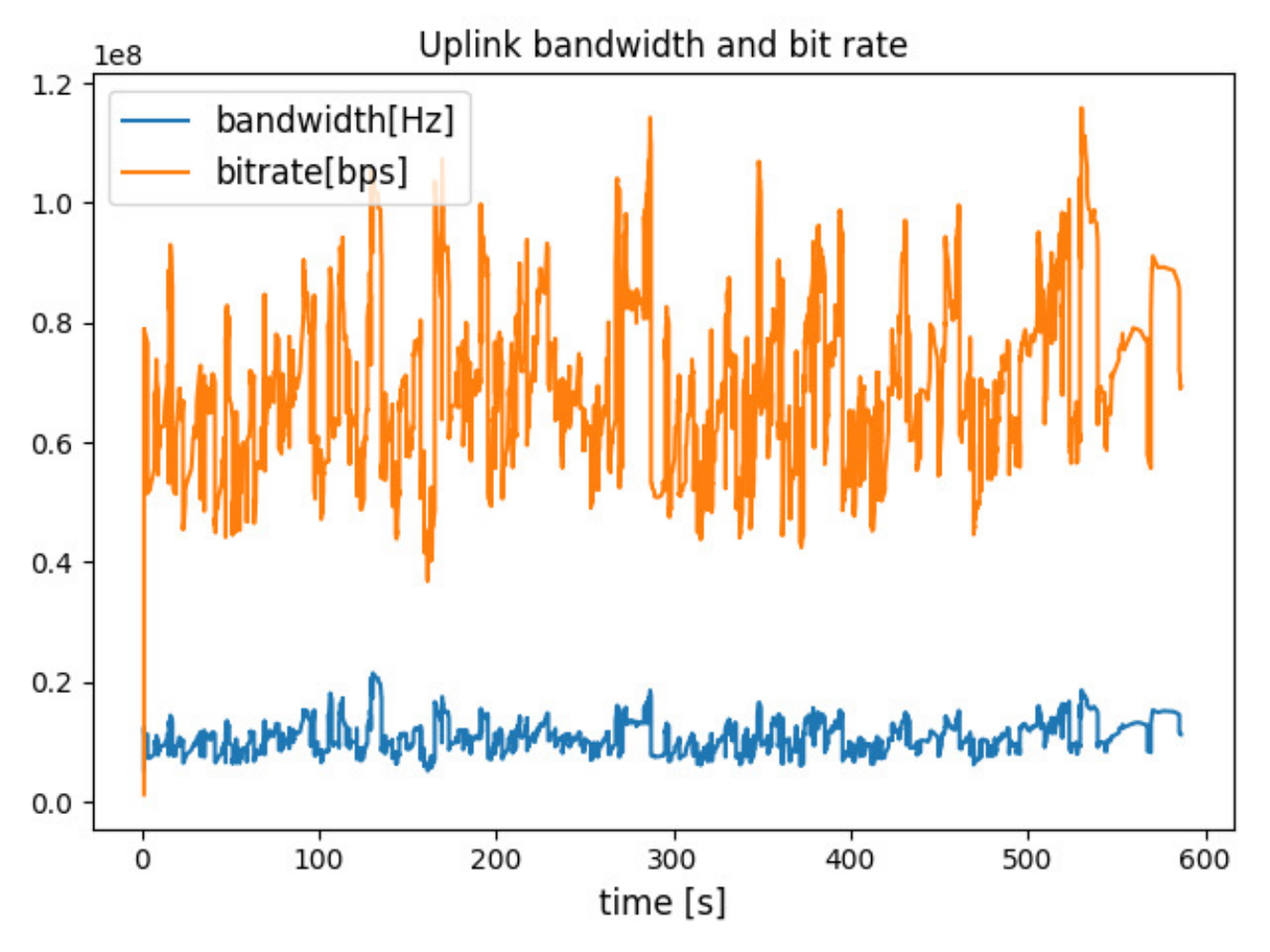}
  \includegraphics[width=0.65\textwidth]{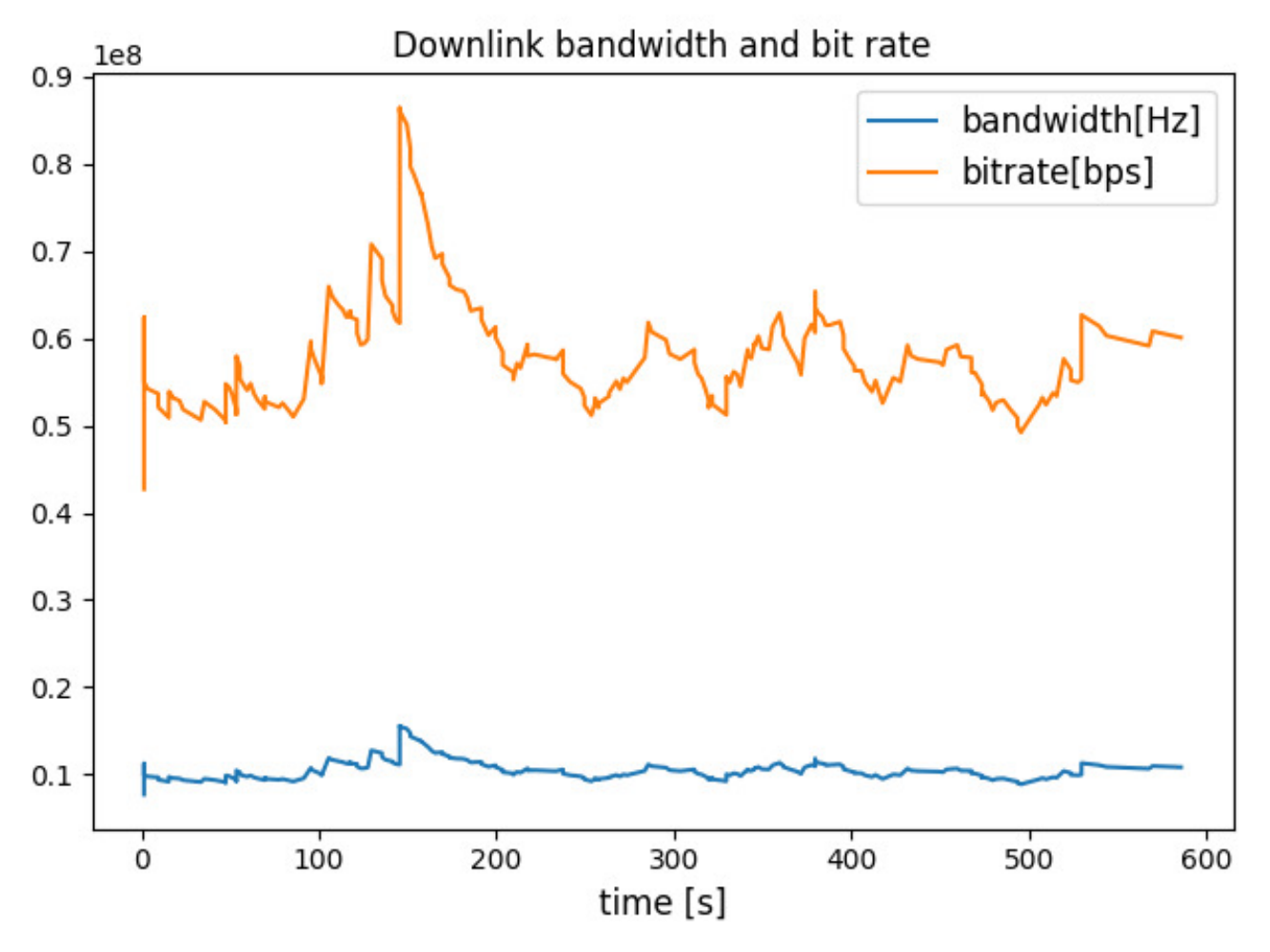}
  \caption{Radio interface bandwidth and bit rate in San Francisco scenario.}
  \label{fig:friscobw}
\end{figure}

After simulating both experiments, it is possible to analyze both outcomes
quantitively in terms of available bandwidth, power consumption and perceived delay.
The only difference between the two experiments resided on the \glspl{EDC}'
dispatching algorithm. However, as the location of both \glspl{AP} and \gls{UE}
devices were the same, each \gls{UE} was connected to the same \glspl{AP} during both
simulations, and the bandwidth share, as well as the applied \gls{MCS}, coincided.
Due to this, the uplink and downlink bandwidth and bit rate perceived by \gls{UE} modules in both experiments is the same, as shown in
Figure~\ref{fig:friscobw}.

\begin{figure}[htb]
  \center
  \includegraphics[width=0.65\textwidth]{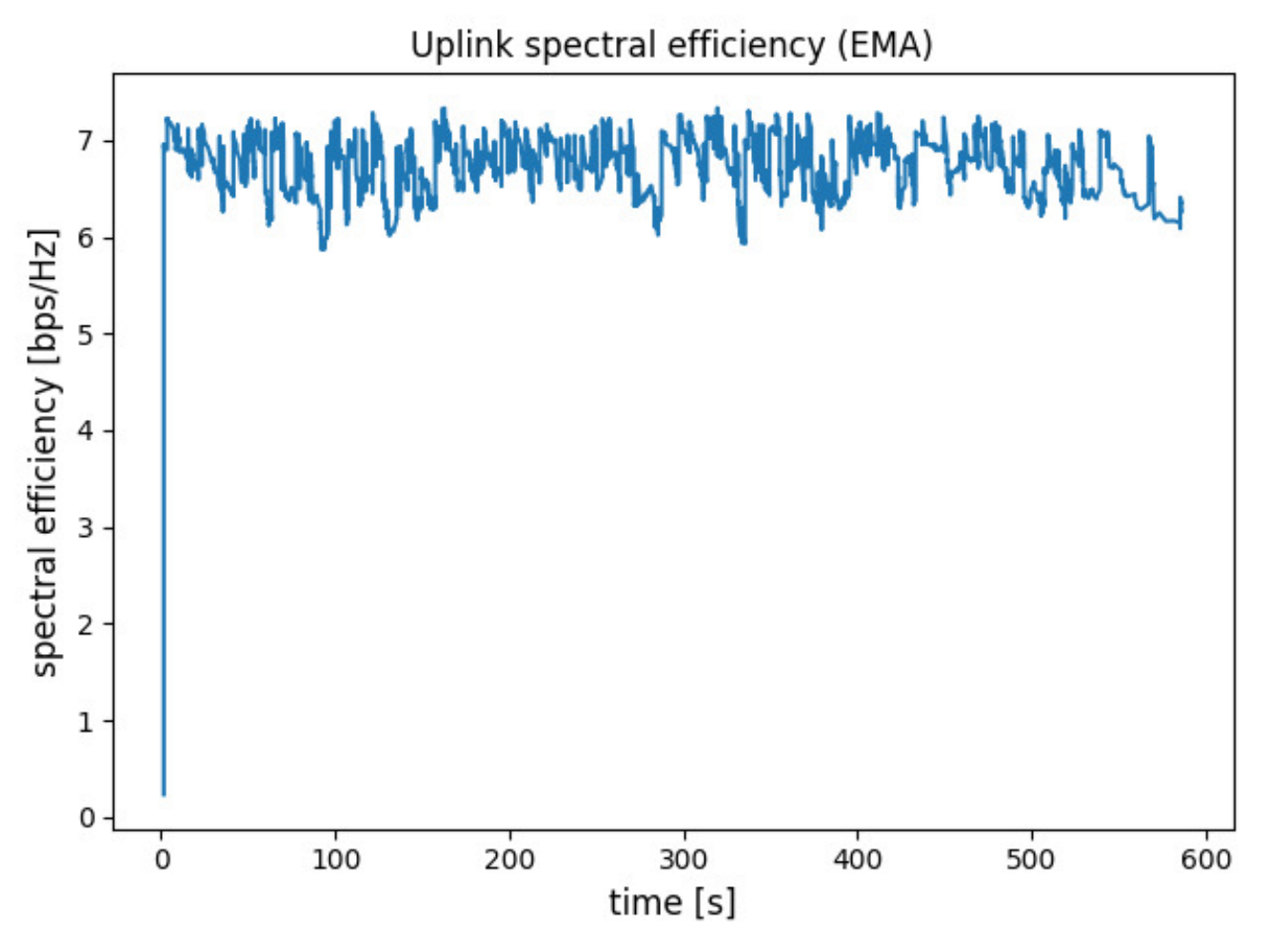}
  \caption{Radio interface uplink spectral efficiency and bit rate in San Francisco scenario.}
  \label{fig:friscoeff}
\end{figure}

In both cases, the bandwidth ranges from 9 to 12 MHz, with a peak  of 16 MHz.
Provided that the scenario has 10 \glspl{AP} and there are 100 taxis, ideally only 10
\gls{UE} devices would be connected to a single \glspl{AP}, thus having 10
MHz - the total available radio bandwidth is 100 MHz per \gls{AP}. Therefore, we can
conclude that the bandwidth share is achieved with Mercury's allocation
manager.
Downlink bandwidth and bit rate, though in different scales, match during all the
simulation. The reason for this is that the downlink spectral efficiency is 5.5547
bps/Hz throughout the simulation. When comparing it to the \gls{MCS} table of the
\gls{NR} interface~\cite{3gpp38214}, it can be seen that it coincides with the
\gls{MCS} with the highest spectral efficiency. As the transmitting power of the \glspl{AP}
is high, and the maximum downlink modulation is 64-\gls{QAM}, the downlink quality is
good enough to provide always the best efficiency. On the other hand, the transmitting
power of the \gls{UE} nodes is not enough for granting the best spectral efficiency.
Moreover, modulations in the uplink may reach 256-\gls{QAM}, a more complex modulation
that requires better \gls{SNR} to be used. Depending on their distance to the
\gls{AP} they are connected to, the used \gls{MCS} varies. Figure~\ref{fig:friscoeff}
shows the uplink spectral efficiency of the proposed scenario.

\begin{figure}[htb!]
	\center
	\includegraphics[width=0.65\textwidth]{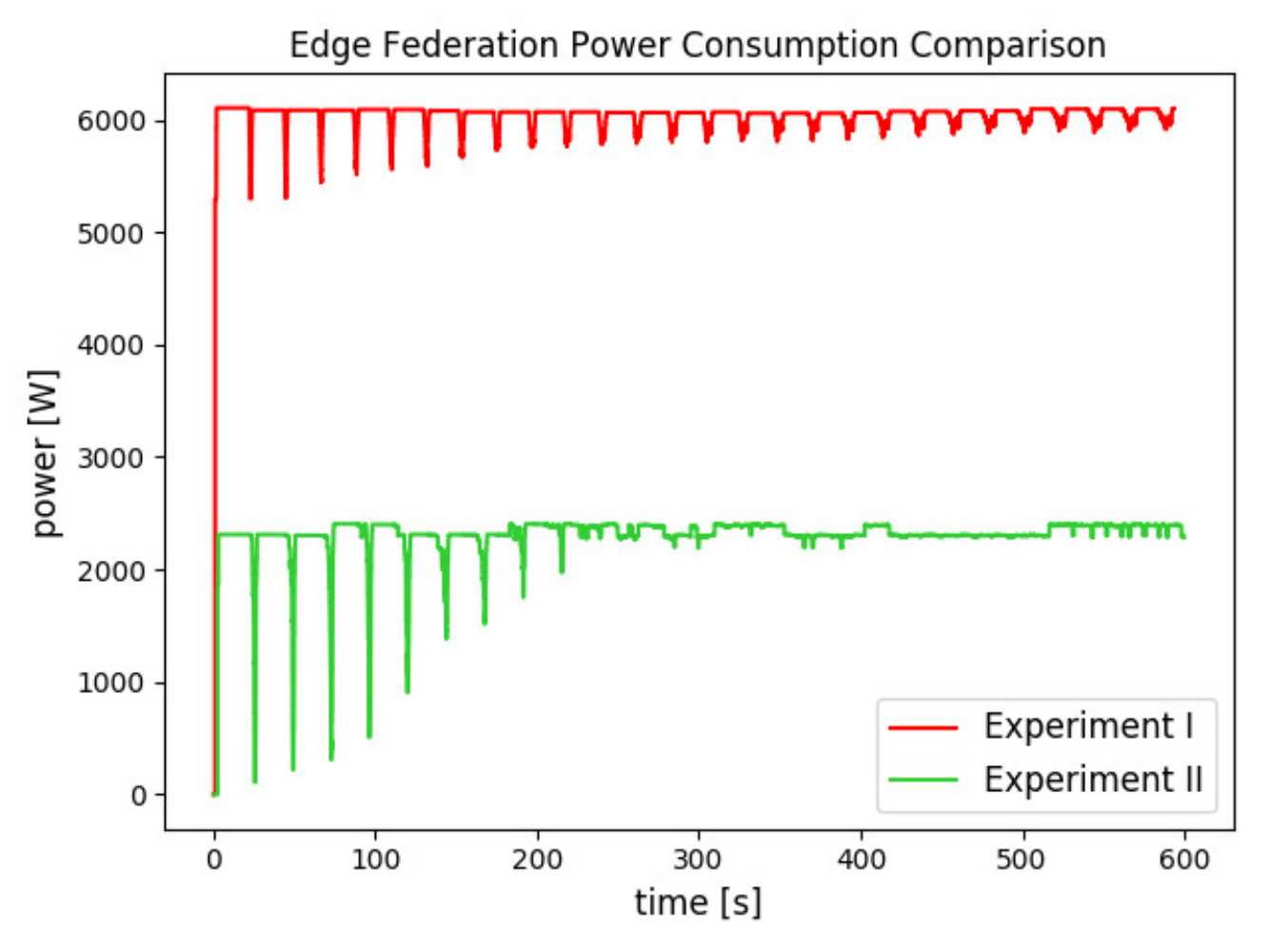}
	\caption{Comparison of EDCs power consumption in San Francisco scenario.}
	\label{fig:friscopower}
\end{figure}

Table~\ref{tab:real_results} gathers different simulation metrics for comparing both
experiments.
Experiment II reported 28\% less power consumption than the first experiment - see
Figure~\ref{fig:friscopower}. This is due to the first experiment's unused
hardware strategy and dispatching algorithm, which keep all the processing units
switched on during the whole simulation. On the other
hand, in experiment II, unused processing units are switched off, and sessions are
dispatched to already switched on processing units if enough resources are available.

\begin{table} [htb]
	\footnotesize
	\centering
	\caption {Real scenario simulation results.}
  	\label{tab:real_results}
	\begin{tabular}{lcc}
  		\toprule
										& Experiment I 	& Experiment II	\\
  		\midrule
		Mean Perceived Delay (ms)		& 35.35 		& 39.31			\\
		Peak Perceived Delay (ms)		& 496.18 		& 2,519.80		\\
		Mean Power Consumption (kW)		& 6.04 			& 2.33   		\\
		Peak Power Consumption (kW)		& 6.10 			& 2.41			\\
  		\bottomrule
  	\end{tabular}
\end{table}

\begin{figure}[htb!]
	\center
	\includegraphics[width=0.65\textwidth]{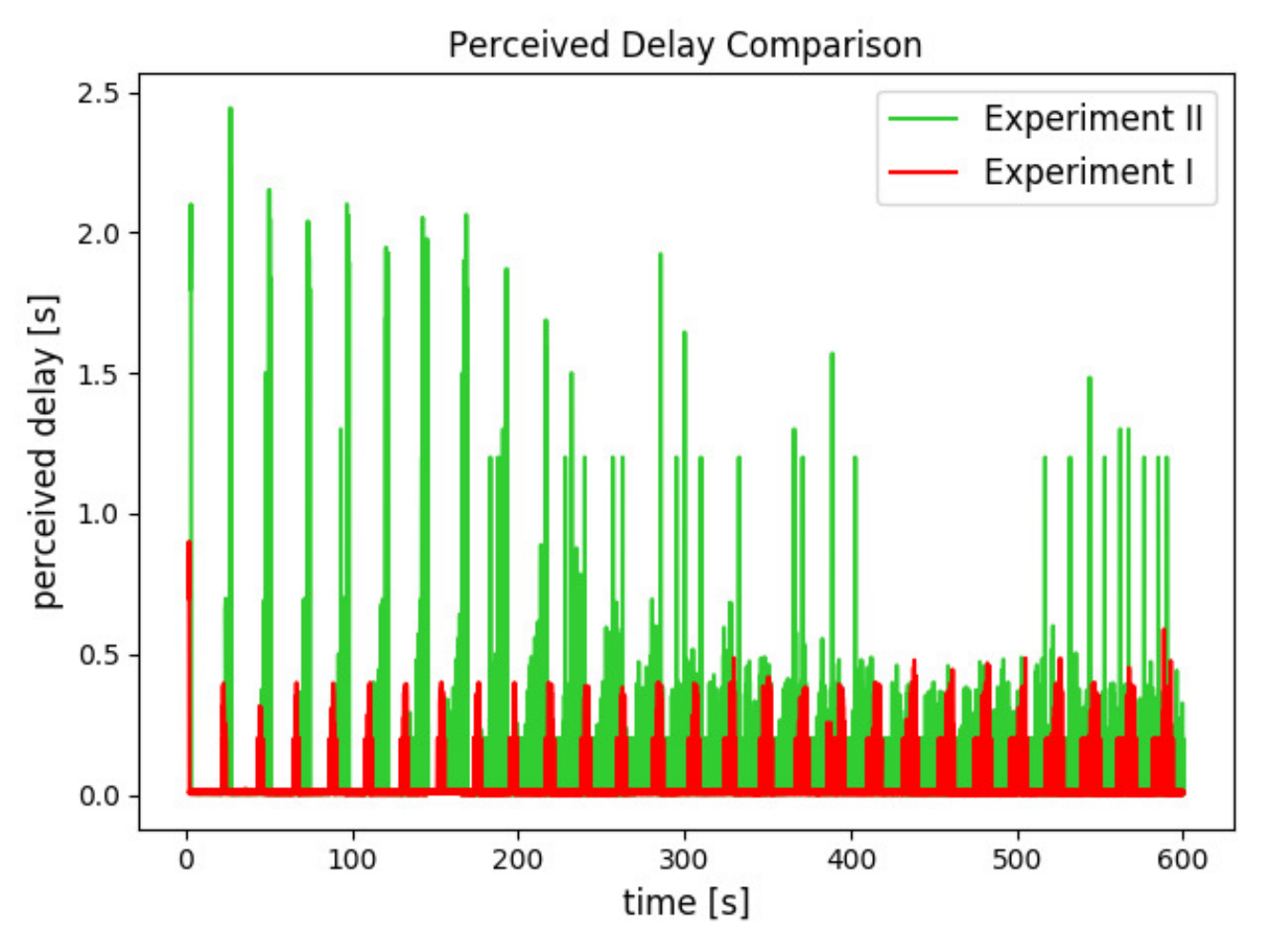}
	\caption{Comparison of delay perceived by UE devices in San Francisco scenario.}
	\label{fig:friscodelay}
\end{figure}

Focusing on the perceived \gls{QoS}, the mean delay perceived in experiment I is 3.96
ms less than in experiment II, which a priori might not seem a significant difference.
However, the second experiment registered a peak delay of 2.52 seconds, in contrast
with the peak delay of 0.50 seconds for the first experiment - i.e., the first
configuration provides a better and more predictable \gls{QoS}.
Figure~\ref{fig:friscodelay} compares the perceived delay in both scenarios.

As shown in this subsection, Mercury includes built-in visualization tools for analyzing during run-time
the scenario outline, bandwidth share, binary rate, perceived delay, and federation power
consumption simulation outcome. Furthermore, Mercury incorporates transducers for
storing any simulation event in \gls{CSV} files in case a more in-depth analysis is
required.
With this output, it is possible to make design decisions depending on the application
under study and its particular technical requirements. Note that the objective of this
paper is to exemplify the workflow of Mercury, as well as its novelties compared to other
available \gls{MS} frameworks.

\subsection{Comparing Mercury With Other Simulators}
To compare Mercury with other alternative fog computing simulation tools, we
defined similar scenarios and analyzed their output. Specifically, we studied
iFogSim~\cite{ifogsim} and EdgeCloudSim~\cite{edgecloudsim}, since these simulators
allow defining generic use cases, as shown in Table~\ref{tab:soa}. All the scenarios were
comprised of 100 \glspl{UE} and three \glspl{EDC}. Other configuration parameters were
dependent on the simulator under study. Since it is not possible to specify exactly the same model of the use case in the different \gls{MS} frameworks, an in-depth numerical comparison between the results obtained by Mercury and by the other \gls{MS} frameworks is out of the scope of this research.

\subsubsection{iFogSim}
iFogSim does not include any spatial conception in its model: all the elements are
interconnected in a tree-based fashion. Communication links are modeled as dedicated
lines between a parent node and each of its children. These links have independent uplink
and downlink bandwidth, as well as a fixed communication delay. Node mobility is not
implemented, and therefore the tree does not change during the simulation. In contrast,
in Mercury, the connections that reduce the perceived latency are selected dynamically
depending on the location of the elements and the instantaneous status of the edge
computing federation. Mercury's behavior coincides with the vision of 5G, in which networks
have cognitive functions that enable them to adapt themselves to the current state of the
infrastructure.

In iFogSim, the user must establish the data flow of the application under study. This
feature enables us to define more complex behaviors based on request-response
communication patterns, which is a useful feature for exploring how fog computing
particularities would affect to a given application. Again, this differs from Mercury, which
is focused on data stream-oriented applications, where the state machine that defines
their behavior is the same for all of them. In Mercury, only timing and data size-related parameters
can be configured.

The power consumption model of \glspl{EDC} in iFogSim consists of idle and active power
consumption. On the other hand, Mercury enables us to define more complex power models in
which processing units within the \glspl{EDC} can be powered  on/off and the
instantaneous power consumption depends on the utilization and
\gls{DVFS} configuration.
Two of the \glspl{EDC} in the scenario had three children \glspl{AP}, while the other
had four instead. Each \gls{AP} had 10 children \glspl{UE}. The wall-clock execution time
of iFogSim was 272 milliseconds, outperforming Mercury's. It is worth to mention that,
with iFogSim, we can neither get nor set information about simulation time.

iFogSim's outcome provided information regarding the average delay and power consumption, but
no real-time progression is shown. As a consequence, we cannot extract the temporal
behavior of our scenario with iFogSim. Gathering data with respect to the temporal
behavior of all the components in the simulation process is necessary to define some
management policies regarding hardware configuration as a function of isolated hot spots
like power consumption peaks, network congestion, etc.

\subsubsection{EdgeCloudSim}
In EdgeCloudSim, \glspl{AP} and \glspl{EDC} are tied together, and each \gls{EDC} defined
by the user implicitly involves an \gls{AP} with its independent \gls{WLAN}. \glspl{EDC}
are interconnected via a \gls{MAN}. In contrast with Mercury, cloud computing is also
part of EdgeCloudSim’s model. All the \glspl{EDC} are connected to the cloud through
a \gls{WAN}. If an \gls{EDC} runs out of computing resources, computing offloading can be
triggered to other \gls{EDC} or to the cloud - at expenses of drastically incrementing
the perceived delay.

EdgeCloudSim enables the user to define a minimum and a maximum number of \glspl{UE}, and
several scenarios are simulated exploring different \gls{UE} densities in this range. Mobility is also
managed internally by the simulator, and therefore it is not possible to define
trajectories. \glspl{UE} are always connected to their closest \gls{AP}, and mobility may
lead to a handover. However, the handover is not modeled in detail: there is no \gls{AP}
discovery process nor radio link quality monitoring. In addition, \gls{UE} location does
not affect the available bandwidth nor its spectral efficiency. In Mercury, the
propagation delay is proportional to the distance between \gls{UE} and their
correspondent \gls{AP}. Besides, in our research, the radio spectrum is shared among \gls{UE} connected
to the same \gls{AP} - i.e., transmission delay varies in simulation time. Furthermore, in Mercury,
the spectral efficiency of the communications may vary depending on the \gls{SNR}.

We defined a scenario with three pairs of \glspl{EDC} and \glspl{AP}. Mobility was enabled
for all the \gls{UE} in the scenario. The simulation time was set to 10 minutes, and it
took one second of execution time, outperforming Mercury's. However, EdgeCloudSim does not
support power consumption models, and this aspect was not explored. Moreover, as in iFogSim,
no temporal data is gathered: EdgeCloudSim provides average measurements about the tasks
executed on \glspl{EDC} and cloud, and about which tasks succeeded and which failed due to either
lack of computing resources or mobility issues.


\subsubsection{Comparison Remarks}

In conclusion, one of iFogSim's most appealing features is the exploration of the effect
of fog computing on a given application. It enables the user to define complex request
response communication sequences and explores how computing offloading can be performed
across the network. iFogSim gives a sense of the mean latency experienced by the
application. This framework also provides a coarse-grained intuition of the power
consumption of the infrastructure. On the other hand, EdgeCloudSim’s strongest point is
that it allows exploring the average \gls{QoS} experienced by a different number of users
depending on the available computing resources. It is useful for a first approach of the
edge computing infrastructure dimensioning depending on the expected demand. However,
information about power consumption is not provided.

Mercury is a framework for fine-grained edge infrastructure dimensioning and operation
tasks. It focuses on data stream-oriented applications. Mercury incorporates optimization tools for
assisting in the most suitable location of edge computing infrastructure. This \gls{MSO}
framework provides a wide range of real-time data - e.g., the usage of the network and the
computing resources, perceived delay, and bandwidth share. It is possible to study the
implications of using different dispatching and hardware management techniques within the
\glspl{EDC} that comprise the edge computing federation. Therefore, Mercury helps to
diminish the impact  of hotspots, leveraging the usage of computing resources in a
location-aware manner. It could impact positively on the operational expenses and the
\gls{QoS} of future fog computing deployments.

\section{Conclusions and Future Work}
\label{sec:6_conclusions}

In this research, we present Mercury, an \gls{MSO} framework to analyze the dimensioning and run-time operation of edge federations in fog computing scenarios. Our framework is designed using \gls{MBSE} methodologies together with the \gls{DEVS} mathematical formalism, providing an explicit separation between the model specification and its corresponding implementation. When defining a complex system, this formalism helps to isolate the detailed specification of its elements in terms of structure, behavior, and their relationships, providing an atomic verification for all of them.

On the one hand, Mercury provides a fine-grained hierarchical model of fog computing environments. The fog computing paradigm provides a solution for distributing the computing infrastructure geographically, closer to data sources. Mercury also includes a set of automatic tools for assisting with the physical location of these infrastructures in the \acrlong{RAN}.

Our model supports location awareness, inherent in fog computing, through a \gls{FaaS}/serverless model that dynamically allocates application service requests. This enables the definition of fine-grained energy efficient strategies due to on-demand resource provisioning. In contrast with other \gls{MS} tools, Mercury is focused on data stream analytics-based services, as fog computing appears as a technology enabler for this type of applications.

\gls{IoT} devices’ mobility  is natively supported by Mercury, which models a federated management of computation offloading that adapts resource allocation to the geographical application demand profile with the aim of optimizing the perceived \gls{QoS}. Inter-module communication models are based on the 5G standard to reflect the impact of networking technologies on the application performance, gathering the effects of bandwidth share, packet loss, perceived delay, and connectivity protocols.

On the other hand, Mercury is implemented on top of the Python 3 distribution of xDEVS, making extensive use of Abstract Base Classes for ensuring high flexibility. This \gls{MSO} framework provides a wide range of simulation real-time data - e.g., the usage of the network and the computing resources, perceived delay, and bandwidth share. These detailed simulation outputs allows the analysis of dynamic optimization strategies in fog computing environments.

Finally, we demonstrate the novelty and major advantages of Mercury, proposing a realistic use case scenario (incorporating real data traces and validated energy consumption models), to show how our \gls{MSO} framework can assist with the dimensioning and operation tasks during the deployment of the edge computing infrastructure. We offer as well a detailed comparison with other state-of-the-art fog computing simulators. 

Mercury will assist with the dimensioning and operation of edge federations prior to actual system deployments, detecting potential bottlenecks while significantly reducing development costs.

\subsection{Future Work}
We are currently working on a new version of Mercury that includes a new
module, the Decision Support System, to explore different configurations for
the scenario elements, for instance, the number of \glspl{EDC}' processing
units and the workload dispatching algorithms. We are also including a cloud
layer that would be connected to the core network, in order to extend
computation offloading. On the other hand, we are designing a new layer to
model \gls{M2M} communications between neighboring \gls{UE} devices aimed to
enhance computation offloading and data sharing at \gls{UE} level. Finally, a
new smart grid layer will be developed to simulate an electrical grid with
different renewable energy sources as well as the energy distribution and
storage systems. Adding these new contributions to Mercury will open the
possibility of studying a wide variety of new optimization issues.

\section*{Acknowledgments}
This project has been partially supported by the Centre for the Development of
Industrial Technology (CDTI) under contracts IDI-20171194, IDI-20171183 and RTC-2017-6090-3
and by the Education and Research Council of the Community of Madrid (Spain), under research grant P2018/TCS-4423.

\section*{References}
\bibliography{bibliographySimpat}

\begin{thebibliography}{10}
\expandafter\ifx\csname url\endcsname\relax
  \def\url#1{\texttt{#1}}\fi
\expandafter\ifx\csname urlprefix\endcsname\relax\def\urlprefix{URL }\fi
\expandafter\ifx\csname href\endcsname\relax
  \def\href#1#2{#2} \def\path#1{#1}\fi

\bibitem{GartnerPR2017}
{Gartner. Press Releases}, {Gartner Says 8.4 Billion Connected "Things" Will Be
  in Use in 2017, Up 31 Percent From 2016} (2017).

\bibitem{deng2016}
R.~Deng, R.~Lu, C.~Lai, T.~H. Luan, H.~Liang, {Optimal Workload Allocation in
  Fog-Cloud Computing Toward Balanced Delay and Power Consumption}, IEEE
  Internet of Things Journal 3~(6) (2016) 1171--1181.
\newblock \href {https://doi.org/10.1109/JIOT.2016.2565516}
  {\path{doi:10.1109/JIOT.2016.2565516}}.

\bibitem{IECedge2017}
{IEC}, {Edge Intelligence}, Tech. rep., International Electrotechnical
  Commission (2017).

\bibitem{cisco2012}
M.~Saad, Fog computing and its role in the internet of things: Concept,
  security and privacy issues, International Journal of Computer Applications
  180 (2018) 7--9.
\newblock \href {https://doi.org/10.5120/ijca2018916829}
  {\path{doi:10.5120/ijca2018916829}}.

\bibitem{shi2016}
W.~Shi, J.~Cao, Q.~Zhang, Y.~Li, L.~Xu, {Edge Computing: Vision and
  Challenges}, IEEE Internet of Things Journal 3~(5) (2016) 637--646.
\newblock \href {https://doi.org/10.1109/JIOT.2016.2579198}
  {\path{doi:10.1109/JIOT.2016.2579198}}.

\bibitem{8089336}
J.~{Pan}, J.~{McElhannon}, Future edge cloud and edge computing for internet of
  things applications, IEEE Internet of Things Journal 5~(1) (2018) 439--449.
\newblock \href {https://doi.org/10.1109/JIOT.2017.2767608}
  {\path{doi:10.1109/JIOT.2017.2767608}}.

\bibitem{yi2015}
S.~{Yi}, Z.~{Hao}, Z.~{Qin}, Q.~{Li}, Fog computing: Platform and applications,
  in: 2015 Third IEEE Workshop on Hot Topics in Web Systems and Technologies
  (HotWeb), IEEE, 2015, pp. 73--78.
\newblock \href {https://doi.org/10.1109/HotWeb.2015.22}
  {\path{doi:10.1109/HotWeb.2015.22}}.

\bibitem{ansys2017}
S.~Sovani, {Fast-Tracking Advanced Driver Assistance Systems (ADAS) and
  Autonomous Vehicles Development with Simulation}, Tech. rep., ANSYS, Inc.
  (2017).

\bibitem{8289317}
G.~{Premsankar}, M.~{Di Francesco}, T.~{Taleb}, Edge computing for the internet
  of things: A case study, IEEE Internet of Things Journal 5~(2) (2018)
  1275--1284.
\newblock \href {https://doi.org/10.1109/JIOT.2018.2805263}
  {\path{doi:10.1109/JIOT.2018.2805263}}.

\bibitem{8360857}
W.~{Li}, T.~{Yang}, F.~C. {Delicato}, P.~F. {Pires}, Z.~{Tari}, S.~U. {Khan},
  A.~Y. {Zomaya}, On enabling sustainable edge computing with renewable energy
  resources, IEEE Communications Magazine 56~(5) (2018) 94--101.
\newblock \href {https://doi.org/10.1109/MCOM.2018.1700888}
  {\path{doi:10.1109/MCOM.2018.1700888}}.

\bibitem{glikson2017}
A.~Glikson, S.~Nastic, S.~Dustdar, Deviceless edge computing: extending
  serverless computing to the edge of the network, 2017, pp. 1--1.
\newblock \href {https://doi.org/10.1145/3078468.3078497}
  {\path{doi:10.1145/3078468.3078497}}.

\bibitem{mercuryrepo}
R.~{C\'{a}rdenas}, {Mercury M\&S\&O Framework for Fog Computing},
  \url{https://github.com/greenlsi/mercury_mso_framework}, [Online; accessed
  6-October-2019].

\bibitem{acceptedSummersim}
R.~{C\'{a}rdenas}, P.~{Arroba}, J.~{Risco-Mart\'{i}n}, J.~{Moya}, Edge
  federation simulator for data stream analytics, in: Proceedings of the 2019
  Summer Simulation Conference, 2019, pp. 1--12.

\bibitem{HeF0WL18}
Y.~He, X.~Fan, F.~Wang, F.~Wang, J.~Liu, Edge computing empowered generative
  adversarial networks for realtime road sensing, in: 26th {IEEE/ACM}
  International Symposium on Quality of Service, IWQoS 2018, Banff, AB, Canada,
  June 4-6, 2018, IEEE/ACM, 2018, pp. 1--2.
\newblock \href {https://doi.org/10.1109/IWQoS.2018.8624148}
  {\path{doi:10.1109/IWQoS.2018.8624148}}.

\bibitem{8270636}
Q.~{Yuan}, H.~{Zhou}, J.~{Li}, Z.~{Liu}, F.~{Yang}, X.~S. {Shen}, Toward
  efficient content delivery for automated driving services: An edge computing
  solution, IEEE Network 32~(1) (2018) 80--86.
\newblock \href {https://doi.org/10.1109/MNET.2018.1700105}
  {\path{doi:10.1109/MNET.2018.1700105}}.

\bibitem{LI2018667}
Y.~Li, A.-C. Orgerie, I.~Rodero, B.~L. Amersho, M.~Parashar, J.-M. Menaud,
  End-to-end energy models for edge cloud-based {IoT} platforms: Application to
  data stream analysis in {IoT}, Future Generation Computer Systems 87 (2018)
  667 -- 678.
\newblock \href {https://doi.org/10.1016/j.future.2017.12.048}
  {\path{doi:10.1016/j.future.2017.12.048}}.

\bibitem{edgecloudsim}
C.~Sonmez, A.~Ozgovde, C.~Ersoy, Edgecloudsim: An environment for performance
  evaluation of edge computing systems, Trans. on Emerging Telecommunications
  Technologies 29~(11) (2018).
\newblock \href {https://doi.org/10.1002/ett.3493}
  {\path{doi:10.1002/ett.3493}}.

\bibitem{ifogsim}
H.~Gupta, A.~Vahid~Dastjerdi, S.~K. Ghosh, R.~Buyya, {iFogSim}: A toolkit for
  modeling and simulation of resource management techniques in the internet of
  things, edge and fog computing environments, Software: Practice and
  Experience 47~(9) (2017) 1275--1296.
\newblock \href {https://doi.org/10.1002/spe.2509}
  {\path{doi:10.1002/spe.2509}}.

\bibitem{6253581}
L.~{Yang}, J.~{Cao}, S.~{Tang}, T.~{Li}, A.~T.~S. {Chan}, A framework for
  partitioning and execution of data stream applications in mobile cloud
  computing, in: 2012 IEEE Fifth International Conference on Cloud Computing,
  2012, pp. 794--802.
\newblock \href {https://doi.org/10.1109/CLOUD.2012.97}
  {\path{doi:10.1109/CLOUD.2012.97}}.

\bibitem{serverlesscon}
{Living in a post-container world: Serverless Architecture Magazine 2019},
  Tech. rep., Serverless Architecture Conference (2019).

\bibitem{macqueen1967}
J.~MacQueen, Some methods for classification and analysis of multivariate
  observations, in: Proceedings of the Fifth Berkeley Symposium on Mathematical
  Statistics and Probability, Volume 1: Statistics, University of California
  Press, Berkeley, Calif., 1967, pp. 281--297.

\bibitem{Zeigler2000}
B.~P. Zeigler, H.~Praehofer, T.~G. Kim, {T}heory of {M}odeling and
  {S}imulation. {I}ntegrating {D}iscrete {E}vent and {C}ontinuous {C}omplex
  {D}ynamic {S}ystems, 2nd Edition, Academic Press, 2000.

\bibitem{RiscoMartin2017}
J.~L. Risco-Martín, S.~Mittal, J.~C. Fabero, M.~Zapater, R.~Hermida,
  Reconsidering the performance of devs modeling and simulation environments
  using the devstone benchmark, SIMULATION 93~(6) (2017) 459--476.

\bibitem{3gpp38214}
{5G; NR; Physical Layer Procedures for Data (3GPP TS 38.214 version 15.2.0
  Release 15)}, Tech. rep., ETSI (2018).

\bibitem{crosshaul}
{5G Crosshaul: the integrated fronthaul/backhaul},
  \url{http://5g-crosshaul.eu}, [Online; accessed 17-June-2019].

\bibitem{nrqualcomm}
{5G NR: 5G New Radio Standard},
  \url{https://www.qualcomm.com/invention/5g/5g-nr}, [Online; accessed
  17-June-2019].

\bibitem{comsnets09piorkowski}
M.~Piorkowski, N.~Sarafijanovoc-Djukic, M.~Grossglauser, {A Parsimonious Model
  of Mobile Partitioned Networks with Clustering}, in: The First International
  Conference on COMmunication Systems and NETworkS (COMSNETS), 2009.

\bibitem{DiazChito201698}
K.~Diaz-Chito, A.~Hern\'{a}ndez-Sabat\'{e}, A.~M. L\'{o}pez, A reduced feature
  set for driver head pose estimation, Applied Soft Computing 45 (2016) 98 --
  107.
\newblock \href {https://doi.org/10.1016/j.asoc.2016.04.027}
  {\path{doi:10.1016/j.asoc.2016.04.027}}.

\bibitem{8057318}
G.~{Ananthanarayanan}, P.~{Bahl}, P.~{Bodí\'{i}k}, K.~{Chintalapudi},
  M.~{Philipose}, L.~{Ravindranath}, S.~{Sinha}, Real-time video analytics: The
  killer app for edge computing, Computer 50~(10) (2017) 58--67.
\newblock \href {https://doi.org/10.1109/MC.2017.3641638}
  {\path{doi:10.1109/MC.2017.3641638}}.

\bibitem{pycharm}
{PyCharm: the Python IDE for Professional Developers},
  \url{https://www.jetbrains.com/pycharm/}, [Online; accessed 22-May-2019].

\end{thebibliography}

\pagebreak
\printglossary[title=Abbreviation List,style=mcolindex,type=main,nonumberlist]

\end{document}